\documentclass[twocolumn,showpacs]
{revtex4}

\usepackage{latexsym}
\usepackage{bm}

\def\tr{{\rm tr}}

\def\ket#1{\mid~\!\!\!{#1}~\!\!\rangle}
\def\bra#1{\langle~\!\!{#1}~\!\!\!\mid}
\def\cH{{\cal H}}
\def\cS{{\cal S}}
\def\cE{{\cal E}}
\def\cU{{\cal U}}

\def\cO{{\cal O}}

\def\IF{if and only if }
\def\QM{quantum mechanics }
\def\qm{quantum mechanics}
\def\cR{{\cal R}}
\def\ON{orthonormal }
\def\${\enskip$}
\def\Sc{Schlosshauer and Fine }
\def\sc{Schlosshauer and Fine}

\begin{document}

\title[ENVARIANCE]{Complete "Born's
rule" from "environment-assisted
invariance"\\
in terms of pure-state twin unitaries}

\author{Fedor Herbut}
\affiliation {Serbian Academy of
Sciences and Arts, Knez Mihajlova 35,
11000 Belgrade, Serbia}

\email{fedorh@mi.sanu.ac.yu}

\date{\today}

\begin{abstract}
Zurek's derivation of the Born rule
from envariance
 (environment-assisted
invariance) is tightened up, somewhat
generalized, and extended to encompass
all possibilities. By this, besides
Zurek's most important work also the
works of 5 other commentators of the
derivation is taken into account, and
selected excerpts commented upon. All
this is done after a detailed theory of
twin unitaries,which are the other face
of envariance.
\end{abstract}

 \maketitle
\rm

\section{INTRODUCTION}

Zurek has introduced \cite{Zurek1} {\it
envariance} (environment-assisted
invariance) in the following way. He
imagined a system \$\cS\$ entangled
with a dynamically decoupled
environment \$\cE\$ altogether
described by a bipartite state vector
\$\ket{\psi}_{\cS \cE}.\$ Further, he
imagined two opposite-subsystem unitary
operators \$u_{\cS}\$ and \$u_{\cE}\$
that "counter-transformed" each other
when elevated to the composite system
\$U_{\cS}\equiv (u_{\cS}\otimes
1_{\cE}),\$ \$U_{\cE}\equiv (1_{\cS}
\otimes u_{\cE}),\$ and applied to the
bipartite state vector, e. g.,
$$U_{\cE}U_{\cS}\ket{\psi}_{\cS
\cE}=\ket{\psi}_{\cS \cE}.\eqno{(1)}$$

Zurek remarked: "When the transformed
property of the system can be so
"untransformed" by acting only on the
environment, it is not the property of
\$\cS."\$ Zurek, further, paraphrases
Bohr's famous dictum: "If the reader
does not find envariance strange, he
has not understood it."

The {\it first aim} of this study is to
acquire a full understanding of
envariance. The wish to understand
envariance as much as possible is not
motivated only by its strangeness, but
also by the fact that Zurek makes use
of it to derive one of the basic laws
of \qm: Born's rule. His argument to
this purpose gave rise to critical
comments and inspired analogous
attempts \cite{Schlossh2},
\cite{Barnum2}, \cite{Mohrhoff},
\cite{Caves}.

Since the term "Born's rule" is not
widely used, the term "probability rule
of \qm" will be utilized instead in
this article.

The probability rule in its general
form states that if \$E\$ is an event
or property (mathematically a projector
in the state space) of the system, and
\$\rho\$ is its state (mathematically a
density operator), then the probability
of the former in the latter is
\$\tr(E\rho ).\$ (This form of the
probability rule is called the "trace
rule"). It is easy to see that an
equivalent, and perhaps more practical,
form of the probability rule is the
following: If \$\ket{\phi}\$ is an
arbitrary state vector of the system,
then \$\bra{\phi}\rho\ket{\phi}\$ is
the probability that in a suitable
measurement on the system in the state
\$\rho\$ the {\it event}
\$\ket{\phi}\bra{\phi}\$ will occur.
This is what is meant by the
probability rule in this article. (For
a proof of the equivalence of the trace
rule and the probability rule of this
article see subsection V.E.) For
brevity, we'll utilize the state vector
\$\ket{\phi}\$ instead of the event
\$\ket{\phi}\bra{\phi}$ throughout.

All derivations of Born's rule from
envariance in the literature are {\it
restricted} to eigen-states
(\$\rho\ket{\phi}=r\ket{\phi},\enskip
r\$ a positive number). Four of the
cited commentators of Zurek's argument
(I have failed to get in touch with
Fine) have pointed out to me that the
restriction can be understood as
natural in the context of (previous)
system-environment interaction, which
has led to decoherence (see
\cite{Schlossh1}, Sec. IIIE4), or if
one takes the relative-state (or
many-worlds) view, where the "observer"
is so entangled with the system in the
measurement that the restriction covers
the general case (cf \cite{Barnum1} and
see the first quotation in subsection
IV.A).

It is the second and {\it basic aim} of
this investigation to follow Zurek's
argument in a general and precise form
using the full power of envariance, and
to complete the argument to obtain the
probability rule, i. e., the formula
\$\bra{\phi}\rho \ket{\phi},\$ beyond
the approach in terms of the Schmidt
decomposition (used in the literature).

In the first subsection of the next
section a precise and detailed
presentation of the Schmidt
decomposition and of its more specific
forms, the canonical Schmidt
decomposition, and the strong Schmidt
decomposition is given. In this last,
most specific form, the antiunitary
correlation operator \$U_a\$, the sole
correlation entity inherent in a given
bipartite state vector (introduced in
previous work \cite{Varenna}) is made
use of. It is the entity that turns the
Schmidt canonical decomposition into
the strong Schmidt decomposition, which
is complete and precise. This entity is
lacking in almost all examples of the
use of the Schmidt decomposition in the
literature. (For an alternative
approach to the correlation operator
via the antilinear operator
representation of bipartite state
vectors see section 2 in \cite{FH06}.)
Twin unitaries, i. e.,
opposite-subsystem unitary operators
that act equally on a given bipartite
state vector, which are hence
equivalent to envariance, are analysed
in detail, and the group of all pairs
of them is derived.

There is another derivation of the full
set of envariance in the recent
literature \cite{Paris}. It is
algebraic, i. e., in terms of matrices
and suitable numbers, whereas the
approach of this study is geometrical,
i. e., it is in terms of state space
decompositions and suitable maps.

In the second subsection of the next
section connection between twin
unitaries and twin Hermitians, i. e.,
so-called twin observables, studied in
detail in pure bipartite states in
previous articles \cite{Varenna},
\cite{DistMeas}, is established. In the
last subsection of the next section a
possibility to extend the notion of
twin unitaries to mixed bipartite
states is shortly discussed. Extension
to twin Hermitians in mixed states was
accomplished in previous work
\cite{saY}.

The second and third subsections of
section II are not necessary for
reading section III, in which,
following Zurek, a complete argument of
obtaining the probability rule is
presented with the help of the group of
all pairs of twin unitaries and
distance in the Hilbert space of linear
Hilbert-Schmidt operators.

In section IV., each of the four
re-derivations of Born's rule from
envariance, and Zurek's most mature
Physical Review article on the subject,
are glossed over and quotations from
them are commented upon from the point
of view of the version presented in
section III.

In concluding remarks of the last
section the main points of this work
are summed up and commented upon.

\section{Mathematical interlude:
strong Schmidt decomposition and twin
unitaries}

The main investigation is in the first
subsection.

\subsection{Pure-state twin unitaries}

We take a completely arbitrary {\it
bipartite state vector}
\$\ket{\Psi}_{12}\$ as given. It is an
arbitrary normalized vector in
\$\cH_1\otimes\cH_2,\$ where the factor
spaces are finite- or
infinite-dimensional complex separable
Hilbert spaces. The statements are, as
a rule, asymmetric in the roles of the
two factor spaces. But, as it is well
known, for every general asymmetric
statement, also its symmetric one,
obtained by exchanging the roles of
\$1\$ and \$2,\$ is valid. We call an
\ON
complete basis simply "basis".\\

The natural framework for the Schmidt
decomposition is {\it general expansion
in a factor-space basis}.

 Let \$\{\ket{m}_1:\forall m\}\$ be an
arbitrary basis in \$\cH_1.\$ Then
there exists a unique expansion
$$\ket{\Psi}_{12}= \sum_m\ket{m}_1
\ket{m}'_2,\eqno{(2a)}$$ where the
generalized expansion coefficients
\$\{\ket{m}'_2: \forall m\}\$ are
elements of the opposite factor space
\$\cH_2,\$ and they depend only on
\$\ket{\Psi}_{12}\$ and the
corresponding basis vectors
\$\ket{m}_1,\$ and not on the entire
basis.

 The generalized expansion coefficients
are evaluated making use of the partial
scalar product: $$\forall m:\quad
\ket{m}'_2=\bra{m}_1\ket{\Psi}_{12}.
\eqno{(2b)}$$

The partial scalar product is evaluated
expanding \$\ket{\Psi}_{12}\$ in
arbitrary bases \$\{\ket{k}_1:\forall
k\}\subset\cH_1,\$
\$\{\ket{l}_2:\forall
l\}\subset\cH_2,\$ and by utilizing the
ordinary scalar products in the
composite and the factor spaces:
$$ \ket{\Psi}_{12}=
\sum_k\sum_l\Big(\bra{k}_1\bra{l}_2
\ket{\Psi}_{12}\Big)\ket{k}_1\ket{l}_2.
\eqno{(2c)}$$ Then (2b) reads $$\forall
m:\quad \ket{m}'_2=\sum_l
\Big(\sum_k\bra{m}_1 \ket{k}_1
\bra{k}_1
\bra{l}_2\ket{\Psi}_{12}\Big)\ket{l}_2,
\eqno{(2d)}$$ and the lhs is
independent of the choice of the bases
in the factor spaces.

Proof is straightforward.\\

Now we define a Schmidt decomposition.
It is well known and much used in the
literature. It is only a springboard
for the theory presented in this
section.

If in the expansion (2a) besides the
basis vectors \$\ket{m}_1\$ also the
"expansion coefficients" \$\ket{m}'_2\$
are orthogonal, then one speaks of a
{\it Schmidt decomposition}. It is
usually written in terms of normalized
second-factor-space vectors
\$\{\ket{m}_2:\forall m\}$:
$$\ket{\Psi}_{12}=\sum_m\alpha_m\ket{m}_1
\ket{m}_2,\eqno{(3a)}$$ where
\$\alpha_m\$ are complex numbers, and
\$\forall m:\enskip\ket{m}_1\$ and
\$\ket{m}_2\$ are referred to as {\it
partners} in a pair of Schmidt states.

The term "Schmidt decomposition" can be
replaced by "Schmidt expansion" or
"Schmidt form". To be consistent and
avoid confusion, we'll stick to the
first term throughout.\\

Expansion (2a) is a {\it Schmidt
decomposition \IF} the
first-factor-space basis \$\{\ket{m}_1:
\forall m\}\$ is an eigen-basis of the
corresponding reduced density operator
\$\rho_1,\$ where
$$\rho_s\equiv\tr_{s'}\Big(\ket{\Psi}_{12}
\bra{\Psi}_{12}\Big),\quad
s,s'=1,2,\quad s\not= s',\eqno{(4)}$$
and \$\tr_s\$ is the partial trace over
\$\cH_s$.\\

Next we define a more specific and more
useful form of the Schmidt
decomposition. It is called canonical
Schmidt decomposition.

The non-trivial phase factors of the
non-zero coefficients \$\alpha_m\$ in
(3a) can be absorbed either in the
basis vectors in \$\cH_1\$ in (3a) or
in those in \$\cH_2\$ (or partly the
former and partly the latter). If in a
Schmidt decomposition (3a) all non-zero
\$\alpha_m\$ are non-negative real
numbers, then we write instead of (3a),
the following decomposition
$$\ket{\Psi}_{12}=\sum_ir_i^{1/2}
\ket{i}_1\ket{i}_2,\eqno{(3b)}$$ and we
confine the sum to non-zero terms (one
is reminded of this by the replacement
of the index \$m\$ by \$i$ in this
notation). Relation (3b) is called a
{\it canonical Schmidt decomposition}.
(The term "canonical" reminds of the
form of (3b), i. e., of \$\forall
i:\enskip r_i^{1/2}>0.\$)

Needless to say that every
\$\ket{\Psi}_{12}\$ can be written as a
canonical Schmidt decomposition.

Each canonical Schmidt decomposition
(3b) is accompanied by the {\it
spectral forms of the reduced density
operators}:
$$\rho_s=\sum_ir_i
\ket{i}_s\bra{i}_s,\quad s=1,2.
\eqno{(5a,b)}$$ (The same eigenvalues
\$r_i\$ appear both in (3b) and in
(5a,b).)

One should note that the topologically
closed ranges \$\bar\cR(\rho_s),\enskip
s=1,2\$ (subspaces) of the reduced
density operators \$\rho_s,\enskip
s=1,2\$ are {\it equally dimensional}.
The range-projectors are $$Q_s=\sum_i
\ket{i}_s\bra{i}_s,\quad s=1,2.
\eqno{(5c,d)}$$ The two reduced density
operators have {\it equal eigenvalues}
\$\{ r_i:\forall i\}\$ (including equal
possible degeneracies).

One has a canonical Schmidt
decomposition (3b) {\it \IF} the
decomposition is bi-orthonormal and all
expansion coefficients are positive.

Proof of these claims is
straightforward.\\

It is high time we introduce {\it the
sole entanglement entity} inherent in
any bipartite state vector, which is
lacking from both forms of Schmidt
decomposition discussed so far. It is
an antiunitary map that takes the
closed range \$\bar\cR(\rho_1)\$ onto
the symmetrical entity
\$\bar\cR(\rho_2).\$ (If the ranges are
finite-dimensional, they are {\it ipso
facto} closed, i. e., they are
subspaces.) The map is called {\it the
correlation operator}, and denoted by
the symbol \$U_a\$ \cite{Varenna},
\cite{DistMeas}.

If a canonical Schmidt decomposition
(3b) is given, then the two \ON bases
of equal power \$\{ \ket{i}_1:\forall
i\}\$ and \$\{\ket{i}_2:\forall i\}\$
define an antiunitary, i. e.,
antilinear and unitary operator
\$U_a,\$ the correlation operator - the
sole correlation entity inherent in the
given state vector \$\ket{\Psi}_{12}$:
$$\forall i:\quad \ket{i}_2\equiv \Big(U_a
\ket{i}_1\Big)_2. \eqno{(6a)}$$

The correlation operator \$U_a,\$
mapping \$\bar\cR(\rho_1)\$ onto
\$\bar\cR(\rho_2),\$ is well defined by
(6a) and by the additional requirements
of antilinearity (complex conjugation
of numbers, coefficients in a linear
combination) and by continuity (if the
bases are infinite). (Both these
requirements follow from that of
antiunitarity.) Preservation of every
scalar product up to complex
conjugation, which, by definition,
makes \$U_a\$ antiunitary, is easily
seen to follow from (6a) and the
requirements of antilinearity and
continuity because \$U_a\$ takes an \ON
basis into another \ON one.\\

Though the canonical Schmidt
decompositions (3b) are non-unique
(even if \$\rho_s,\enskip s=1,2\$ are
non-degenerate in their positive
eigenvalues, there is the
non-uniqueness of the phase factors of
\$\ket{i}_1\$), the correlation
operator \$U_a\$ is {\it uniquely}
implied by a given bipartite state
vector \$\ket{\Psi}_{12}$.

This claim is proved in Appendix A.

The uniqueness of \$U_a\$ when
\$\ket{\Psi}_{12}\$ is given is a
slight compensation for the trouble one
has treating an antilinear operator.
(Though the difficulty is more
psychological than practical, because
all that distinguishes an antiunitary
operator from a unitary one is its
antilinearity - it complex-conjugates
the numbers in any linear combination -
and its property that it preserves the
absolute value, but complex-conjugates
every scalar product.) The full
compensation comes from the usefulness
of \$U_a$.

Once the \ON bases
\$\{\ket{i}_1:\forall i\}\$ and
\$\{\ket{i}_2:\forall i\}\$ of a
canonical Schmidt decomposition (3b)
are given, one can write $$U_a=
\sum_i\ket{i}_2K\bra{i}_1,\eqno{(6b)}$$
where \$K\$ denotes complex
conjugation. For instance, $$U_a\ket{
\phi}_1=\sum_i(\bra{i}_1\ket{\phi}_1)
^*\ket{i}_2.\eqno{(6c)}$$\\

We finally introduce the most specific
form of Schmidt decomposition. We call
it a strong Schmidt decomposition.

If one rewrites (3b) in terms of the
correlation operator by substituting
(6a) in (3b), then it takes the form
$$\ket{\Psi}_{12}=\sum_ir_i^{1/2}
\ket{i}_1\Big(U_a\ket{i}_1\Big)_2.
\eqno{(3c)}$$ This is called a
{\it strong Schmidt decomposition}.\\

If a strong Schmidt decomposition (3c)
is written down, then it can be viewed
in two opposite ways:

(i) as a given bipartite state vector
\$\ket{\Psi}_{12}\$ defining its two
inherent entities, the reduced density
operator \$\rho_1\$ in spectral form
(cf (5a)) and the correlation operator
\$U_a\$ (cf (6a)), both relevant for
the entanglement in the state vector;
and

(ii) as a given pair \$(\rho_1,U_a)\$
(\$U_a\$ mapping antiunitarily
\$\bar\cR(\rho_1)\$ onto some equally
dimensional subspace of \$\cH_2\$)
defining a bipartite state vector
\$\ket{\Psi}_{12}$.

The second view of the strong Schmidt
decomposition allows a systematic
generation or classification of all
state vectors in \$\cH_1\otimes\cH_2\$
(cf \cite{gener}).\\

One has $$\rho_2=U_a\rho_1U_a^{-1}Q_2,
\quad \rho_1=U_a^{-1}\rho_2U_aQ_1
\eqno{(7a,b)}$$ (cf (6a) and (5a,b)).
Thus, the reduced density operators
are, essentially, "images" of each
other via the correlation operator.
(The term "essentially" points to the
fact that the dimensions of the null
spaces are independent of each other.)
This property is called {\it twin
operators}.

When one takes into account the {\it
eigen-subspaces} \$\cR(Q_s^j)\$ of
\$\rho_s\$ corresponding to (the
common) distinct positive eigenvalues
\$r_j\$ of \$\rho_s,\$ where \$Q_s^j\$
projects onto the
\$r_j-$eigen-subspace, \$s=1,2,\$ then
one obtains a {\it geometrical view} of
the {\it entanglement} in a given state
\$\ket{\Psi}_{12}\$ in terms of the
so-called {\it correlated subsystem
picture} \cite{Varenna}:
$$\bar\cR(\rho_s)=\sum_j^{\oplus}\cR(Q_s
^j),\quad s=1,2,\eqno{(7c,d)}$$ where
\$"\oplus"\$ denotes an orthogonal sum
of subspaces,
$$\forall j:\quad
\cR(Q_2^j)=U_a\cR(Q_1^j),\quad
\cR(Q_1^j)=U_a^{-1}\cR(Q_2^j),
\eqno{(7e,f)}$$ and, of course,
$$\bar\cR(\rho_2)=U_a\bar\cR(\rho_1),
\quad\bar\cR(\rho_1)=U_a^{-1}
\bar\cR(\rho_2).\eqno{(7g,h)}$$

In words, the correlation operator
makes not only the ranges of the
reduced density operators "images" of
each other, but also the
positive-eigenvalue eigen-subspaces.
Equivalently, the correlation operator
makes the eigen-decompositions of the
ranges "images" of each other.

One should note that all
positive-eigenvalue eigen-subspaces
\$\cR(Q_s^j)\$ are finite dimensional
because \$\sum_ir_i=1\$ (a consequence
of the normalization of
\$\ket{\Psi}_{12}\$), and hence no
positive-eigenvalue can have infinite
degeneracy.

The correlated subsystem picture of a
given bipartite state vector is very
useful in investigating remote
influences (as a way to understand
physically the entanglement in the
composite state) (see \cite{DistMeas},
and \cite{FH06}).

We will need the correlated subsystem
picture of \$\ket{\Psi}_{12}\$ for the
basic result of this section given
below: the second theorem on twin
unitaries. Namely, we now introduce
this term for the pairs \$(U_1,U_2)\$
following a long line of research on
analogous Hermitian operators (see the
last mentioned references and the next
subsection).\\

If one has two opposite factor-space
unitaries \$u_1\$ and \$u_2\$ that, on
defining \$U_1\equiv (u_1\otimes 1_2)\$
and \$U_2\equiv (1_1\otimes u_2),\$
{\it act equally} on the given
composite state vector
$$U_1\ket{\Psi}_{12}=U_2\ket{\Psi}_{12},
\eqno{(8a)}$$ then one speaks of {\it
twin unitaries} (unitary twin
operators). They give another,
equivalent, view of envariance (see the
Introduction), since, rewriting (8a) as
$$U_2^{-1}U_1\ket{\Psi}_{12}=
\ket{\Psi}_{12},\eqno{(8b)}$$ one can
see that \$U_2^{-1}\$ "untransforms"
the action of \$U_1\$ (cf (1)).

It is easy to see that \$U_1
\ket{\Psi}_{12}\bra{\Psi}_{12}U_1^{-1}=
U_2
\ket{\Psi}_{12}\bra{\Psi}_{12}U_2^{-1}\$
is equivalent to $$U_1\ket{\Psi}_{12}=
e^{i\lambda}U_2\ket{\Psi}_{12},
\eqno{(8c)}$$ where \$\lambda\in${\bf
R}$_1.\$ This does not diminish the
usefulness of definition (8a), because,
if (8c) is valid for a pair
\$(U_1,U_2),\$ then one only has to
replace these operators by
\$(U_1,e^{i\lambda}U_2) \$, and the
latter satisfy (8a).

Henceforth, we will write \$U_s\$ both
for \$u_s,\enskip s=1,2,\$ and for
\$(1_1\otimes u_2)\$ or \$(u_1\otimes
1_2)\$ (cf (1)).\\

{\bf First Theorem on twin unitaries.}
Opposite factor-space unitaries \$U_1\$
and \$U_2\$ are twin unitaries {\it
\IF} the following two conditions are
satisfied:

(i) they are symmetry operators of the
corresponding density operators:
$$U_s\rho_sU_s^{-1}=\rho_s,\quad s=1,2,
\eqno{(8d,e)}$$ and

(ii) they are the correlation-operator
"images" of each other's inverse.
Writing \$Q_s^{\perp}\equiv
1_s-Q_s,\enskip s=1,2,\$ this reads:
$$U_2=U_aU_1^{-1}U_a^{-1}Q_2+U_2Q_2
^{\perp},\eqno{(8f)}$$ $$
U_1=U_a^{-1}U_2^{-1}U_aQ_1+U_1Q_1^{\perp}.
\eqno{(8g)}$$ (The second terms on the
rhs of (8f) and (8g) mean that \$U_s\$
is arbitrary in the null space
\$\cR(Q_s ^{\perp})\$ of
\$\rho_s,\enskip
s=1,2.\$)\\

{\it Proof. Necessity.} $$U_1\rho_1=
U_1\tr_2\Big(\ket{\Psi}_{12}
\bra{\Psi}_{12}\Big)=$$ $$\tr_2\Big(
U_1\ket{\Psi}_{12}\bra{\Psi}_{12}\Big)=
\tr_2\Big((U_2\ket{\Psi}_{12})
\bra{\Psi}_{12}\Big)=$$ $$\tr_2\Big(
(\ket{\Psi}_{12}\bra{\Psi}_{12})U_2\Big)
=\tr_2\Big(\ket{\Psi}_{12}
\bra{\Psi}_{12}U_1\Big)=\rho_1U_1.$$
Symmetrically one derives (8e).

Applying the definition of twin
unitaries in the envariance form (8b)
to \$\ket{\Psi}_{12},\$ written as a
strong Schmidt decomposition (3c), one
obtains $$\sum_ir_i^{1/2}\Big(U_1
\ket{i}_1\Big)U_2^{-1}\Big(U_a\ket{i}_1
\Big)_2=\sum_ir_i^{1/2}\ket{i}_1
\Big(U_a\ket{i}_1\Big)_2.$$ On account
of the unitary property of \$U_1\$ and
\$U_2^{-1},\$ the lhs is
bi-orthonormal, hence also
\$\{U_1\ket{i}_1:\forall i\}\$ is an
eigen-basis of \$\rho_1\$ in
\$\bar\cR(\rho_1)\$ due to the
necessary and sufficient condition for
a Schmidt decomposition (see above
(4)). Then, one can rewrite the lhs as
the strong Schmidt decomposition with
this basis. Thus, one obtains
$$\sum_ir_i^{1/2}\Big(U_1\ket{i}_1\Big)
U_2^{-1}\Big(U_a\ket{i}_1\Big)_2=$$ $$
\sum_ir_i^{1/2}\Big(U_1\ket{i}_1\Big)
\Big(U_aU_1\ket{i}_1\Big)_2.$$ Since
the generalized expansion coefficients
are unique, one concludes
$$U_2^{-1}U_aQ_1= U_aU_1Q_1$$ (cf (5c)).
One has \$U_1=U_1Q_1+ U_1Q_1^{\perp}\$
as a consequence of relation (8d),
which has been proved already, and
which implies commutation with all
eigen-projectors \$Q_1^j,\$ and hence
also with \$Q_1=\sum_jQ_1^j\$ (cf
(7c)). Therefore, the obtained relation
amounts to the same as (8g). The
symmetrical argument establishes (8f).
(Note that here one starts with the
decomposition that is symmetrical to
(3c), in which an eigen-sub-basis of
\$\rho_2\$ is chosen spanning
\$\bar\cR(\rho_2),\$ and \$U_a\$ is
replaced by \$U_a^{-1}.\$)

{\it Sufficiency.} Assuming validity of
(8d), it immediately follows that
besides \$\{\ket{i}_1:\forall i\}\$ (cf
(3c)) also \$\{U_1\ket{i}_1:\forall
i\}\$ is an eigen-sub-basis of
\$\rho_1\$ spanning
\$\bar\cR(\rho_1).\$ Hence, we can
write a strong Schmidt decomposition as
follows:
$$\ket{\Psi}_{12}=\sum_i\Big(U_1\ket{i}_1
\Big)\Big(U_aU_1\ket{i}_1\Big)_2.$$
Substituting here (8g) in the second
factors,
$$\ket{\Psi}_{12}=\sum_i\Big(U_1\ket{i}_1
\Big)\Big(U_2^{-1}U_a\ket{i}_1\Big)_2$$
ensues. In view of the strong Schmidt
decomposition (3c), this amounts to
\$\ket{\Psi}_{12}=U_1U_2^{-1}
\ket{\Psi}_{12},\$ i. e., (8b), which
is equivalent to (8a), is
obtained.\hfill $\Box$\\

It is straightforward to show (along
the lines of the proof just presented)
that the twin unitaries are also
responsible for the non-uniqueness of
strong (or of canonical) Schmidt
decomposition. To put this more
precisely, besides (3c) (besides (3b))
all other strong Schmidt decompositions
(canonical Schmidt decompositions) are
obtained by replacing
\$\{\ket{i}_1:\forall i\}\$ in (3c) by
\$\{U_1\ket{i}_1:\forall i\},\$ where
\$[U_1,\rho_1]=0\$ (by replacing
\$\{\ket{i}_1\ket{i}_2:\forall i\}\$ in
(3b) by \$\{\Big(U_1\ket{i}_1\Big)
\Big(U_2^{-1}\ket{i}_2\Big):\forall
i\},\$ where \$[U_s,\rho_s]=0,\enskip
s=1,2,\$and (8f) is satisfied).\\

The set of all pairs of twin unitaries
\$(U_1,U_2)\$ is a {\it group}, if one
defines the composition law by
\$(U_1',U_2')\times (U_1,U_2)\equiv
(U_1'U_1,U_2U_2')\$ (note the inverted
order in \$\cH_2\$), and taking the
inverse turns out to be
\$(U_1,U_2)^{-1}=
(U_1^{-1},U_2^{-1}).\$ This claim is
proved in Appendix B.

Having in mind the subsystem picture
(7a)-(7h) of \$\ket{\Psi}_{12},\$ it is
immediately seen that the first theorem
on twin unitaries can be cast in the
following equivalent form.\\

{\bf Second Theorem on twin unitaries.}
The group of {\it all} twin unitaries
\$(U_1,U_2)\$ consists of {\it all}
pairs of opposite factor-space
unitaries that reduce in every
positive-eigenvalue eigen-subspace
\$\cR(Q_s^j),\enskip s=1,2\$ (cf
(7c,d)), and the reducees are connected
by relations (8f,g) {\it mutatis
mutandis}, or, equivalently, by (8f,g)
in which \$Q_s\$ is replaced by
\$Q_s^j,\enskip s=1,2,\$ and this is
valid simultaneously for all
\$j-$components.

In the language of formulae, we have
{\it all} pairs of unitaries
\$(U_1,U_2)\$ that can be written in
the form
$$U_s=\sum_jU_s^jQ_s^j+
U_sQ_s^{\perp},\quad
s=1,2,\eqno{(9a,b)}$$ $$\forall j:\quad
U_2^jQ_2^j=U_a(U_1^j)^{-1}U_a^{-1}Q_2^j,
\eqno{(9c)}$$ $$
U_1^jQ_1^j=U_a^{-1}(U_2^j)^{-1}U_aQ_1^j.
\eqno{(9d)}$$

Note that within each
positive-eigenvalue subspace
\$\cR(Q_s^j)\$ of \$\rho_s,\enskip
s=1,2,\$ {\it all} unitaries are
encompassed (but not independently, cf
(9c,d)). This will be important in the
application in the next section.

The next two (short) subsections round
out the study of twin unitaries. The
reader who is primarily interested in
the argument leading to the probability
rule
is advised to skip them.\\

\subsection{Connection with twin
Hermitians}

There is a notion closely connected
with twin unitaries in a pure bipartite
state: it is that of twin Hermitians
(in that state). If a pair
\$(H_1,H_2)\$ of opposite factor-space
Hermitian operators commute with the
corresponding reduced density
operators, and
$$H_2=U_aH_1U_a^{-1}Q_2+H_2Q_2^{\perp},
\quad H_1=U_a^{-1}H_2U_aQ_1
+H_1Q_1^{\perp} \eqno{(10a,b)}$$ is
valid then one speaks of twin Hermitian
operators. (Relations (10a,b), in
analogy with (8f,g), state that the
reducees in the ranges of the reduced
density operators are "images" of each
other, and the reducees in the null
spaces are completely arbitrary.)

One should note that twin unitaries
are, actually, defined analogously. To
see this, one has to replace
\$U_s^{-1}\$ by \$U_s^{\dag}\$ in
(8f,g), and \$H_s\$ by
\$H_s^{\dag},\enskip s=1,2,\$ in
(10a,b).

Twin Hermitians have important physical
meaning \cite{DistMeas}, \cite{FH06}.
But here we are only concerned with
their connection with twin unitaries.

If \$U_s,\enskip s=1\$ or \$s=2\$ are
symmetry operators of the corresponding
reduced density operators, i. e., if
they commute, then there exist
Hermitian operators that also commute
with the latter and
$$U_s=e^{iH_s}Q_s+U_sQ_s^{\perp},
\enskip s=1\enskip\mbox{or}\enskip
s=2\eqno{(11a,b)}$$ is valid. And {\it
vice versa}, if \$H_s,\enskip s=1\$ or
\$s=2\$ are Hermitians that commute
with the corresponding reduced density
operators, then there exist analogous
unitaries given by (11a,b). (The
unitary and Hermitian reducees in the
ranges determine each other in (11a,b),
and the reducees in the null spaces are
arbitrary.)

The latter claim is obvious. But to see
that also the former is valid, one
should take into account that
commutation with the corresponding
reduced density operator implies
reduction in each (finite dimensional)
positive-eigenvalue eigen-subspace (cf
(7c,d)). Then one can take the spectral
form of each reducee of \$U_s\$, and
(11a,b) becomes obvious (and the
corresponding reducees of \$H_s\$ are
unique if their eigenvalues are
required to be, e. g., in the intervals
\$[0,2\pi).\$)

The connection (11a,b), which goes in
both directions, can be extended to
twin operators.

If \$(U_1,U_2)\$ are twin unitaries,
then (11a,b) (with "or" replaced by
"and") determine corresponding twin
Hermitians, and {\it vice versa}, if
\$(H_1,H_2)\$ are twin Hermitians, then
the same relations determine
corresponding twin unitaries.\\

\subsection{Mixed states}

If \$\rho_{12}\$ is a {\it mixed
bipartite density operator}, then we no
longer have the correlation operator
\$U_a\$ and the correlated subsystem
picture (7a)-(7h). Nevertheless, in
some cases twin Hermitians, defined by
$$H_1\rho_{12}=H_2\rho_{12}
\eqno{(12a,b)}$$ have been found
\cite{saY}. (Their physical meaning was
analogous to that in the pure-state
case.) It was shown that (12a,b)
implied
$$[H_s,\rho_s]=0,\quad s=1,2,
\eqno{(12c,d)}$$ where \$\rho_s\$ are
again the reduced density operators.
(Unlike in the case when \$\rho_{12}\$
is a pure state, in the mixed-state
case the commutations (12c,d) are not
sufficient for possessing a twin
operator.)

Relations (12c,d), in turn, again imply
reduction of \$H_s\$ in every
positive-eigenvalue eigen-subspace
\$\cR(Q_s^j)\$ of \$\rho_s,\enskip
s=1,2,\$ but now the dimensions of the
corresponding, i. e., equal-j,
eigen-subspaces are, unlike in (7c,d),
completely independent of each other
(but finite dimensional). In each of
them, relations (11a,b) (with "and"
instead of "or") hold true, and define
{\it twin unitaries} satisfying (8a)
with \$\rho_{12}\$ instead of
\$\ket{\Psi}_{12}$.

Thus, in some cases, the concept of
envariance can be extended to mixed
states.

\section{BORN'S RULE FROM TWIN UNITARIES}

The forthcoming argument is given in 5
stages; the first 3 stages are an
attempt to tighten up and make more
explicit, Zurek's argument
\cite{Zurek1}, \cite{Zurek2},
\cite{Zurek3}, \cite{Zurek4}
by
somewhat changing the approach, and
utilizing the group of all pairs of
twin unitaries (presented in the first
subsection of the preceding section).
The change that is introduced is,
actually, a generalization. Zurek's
"environment", which, after the
standard interaction with the system
under consideration, establishes
special, measurement-like correlations
with it, is replaced. Instead, an
entangled bipartite pure state
\$\ket{\Psi}_{12}\$ is taken, where
subsystem \$1\$ is the system under
consideration, and \$2\$ is some
opposite subsystem with an {\it
infinite dimensional} state space
\$\cH_2.\$  We shall try to see to what
extent and how the quantum probability
rule follows from the quantum
correlations, i. e., the entanglement
in \$\ket{\Psi}_{12}$.

The forth stage is new. It is meant to
extend the argument to states
\$\ket{\phi}_1\$ which are not
eigenvectors of the reduced density
operator \$\rho_1\equiv\tr_2\Big(
\ket{\Psi}_{12}\bra{\Psi}_{12}\Big).\$
The fifth stage is also new. It extends
the argument to isolated (not
correlated) systems.

Let \$\ket{\Psi}_{12}\$ be an arbitrary
entangled bipartite state vector. We
assume that subsystems \$1\$ and \$2\$
are not interacting. (They may have
interacted in the past and thus have
created the entanglement. But it also
may have been created in some other
way; e. g., by an external field as the
spatial-spin entanglement in a
Stern-Gerlach apparatus.)

We want to obtain the probability rule
in subsystem \$1.\$ By this we assume
that there exist probabilities, and we
do not investigate why this is so; we
only want to obtain their form.\\

The FIRST STIPULATION is: {\it (a)}
Though the given pure state
\$\ket{\Psi}_{12}\$ determines all
properties in the composite system,
therefore also all those  of subsystem
\$1,\$ the latter must be {\it
determined actually by the subsystem
alone}. This is, by (vague) definition,
what is meant by {\it local}
properties.

{\it (b) There exist local or subsystem
probabilities} of all elementary events
\$\ket{\phi}_1\bra{\phi}_1,\$
\$\ket{\phi}_1\in\cH_1.\$ (As it has
been stated, we will write the event
shortly as the state vector that
determines it.)

Since
\$\ket{\Psi}_{12}\in\Big(\cH_1\otimes
\cH_2\Big),\$ subsystem \$1\$ is
somehow connected with the state space
\$\cH_1,\$ but it is not immediately
clear precisely how. Namely, since we
start out {\it without the probability
rule}, the reduced density operator
\$\rho_1\equiv\tr_2\Big(\ket{\Psi}_{12}
\bra{\Psi}_{12}\Big),\$ though
mathematically at our disposal, is yet
devoid of physical meaning. We need a
precise definition of what is local or
what is the subsystem state. We will
achieve this gradually, and thus
\$\rho_1\$ {\it will be gradually
endowed with the standard physical
meaning}.\\

The SECOND STIPULATION is that
subsystem or {\it local properties must
not be changeable by remote action}, i.
e., by applying a second-subsystem
unitary \$U_2\$ to \$\ket{\Psi}_{12}\$
or any unitary \$U_{23}\$ applied to
the opposite subsystem with an ancilla
(subsystem \$3\$).

If this were not so, then there would
be no sense in calling the properties
at issue "local" and not "global" in
the composite state. We are dealing
with a {\it definition of local} or
subsystem properties. By the first
stipulation, the probability rule that
we are endeavoring to obtain should be
local.

The most important part of the precise
mathematical formulation of the second
stipulation is in terms of twin
unitaries (cf (8a)). No local unitary
\$U_1\$ that has a twin \$U_2\$ must be
able to change any local property.\\

{\bf Stage one.} We know from the First
Theorem on twin unitaries that such
local unitaries \$U_1\$ are all those
that commute with \$\rho_1\$ (cf (8d))
and no others. In this way the
mathematical entity \$\rho_1\$ is
already beginning to obtain some
physical relevance for local
properties.

We know from the Second Theorem on twin
unitaries that we are dealing with
\$U_1\$ that are orthogonal sums of
{\it arbitrary} unitaries acting within
the positive-eigenvalue eigen-subspaces
of \$\rho_1\$ (cf (9a)).

Let \$\ket{\phi}_1\$ and
\$\ket{\phi}'_1\$ be any two distinct
state vectors from one and the same
positive-eigenvalue eigen-subspace
\$\cR(Q_1^j)\$ of \$\rho_1.\$
Evidently, there exists a unitary
\$U_1^j\$ in this subspace that maps
\$\ket{\phi}_1\$ into
\$\ket{\phi}'_1,\$ and, adding to it
orthogonally any other eigen-subspace
unitaries (cf (9a)), one obtains a
unitary \$U_1\$ in \$\cH_1\$ that has a
twin, i. e., the action of which can be
given rise to from the remote second
subsystem. ("Remote" here refers in a
figurative way to lack of interaction.
Or, to use Zurek's terms, \$1\$ and
\$2\$ are assumed to be "dynamically
decoupled" and "causally
disconnected".) Thus, we conclude that
the two first-subsystem states at issue
must have the {\it same probability}.

In other words, arguing {\it ab
contrario}, if the probabilities of the
two distinct states were distinct,
then, by remote action (by applying the
twin unitary \$U_2\$ of the above
unitary \$U_1\$ to
\$\ket{\Psi}_{12}\$), one could
transform one of the states into the
other, which would locally mean
changing the probability value without
any local cause.

Putting our conclusion differently, all
eigen-vectors of \$\rho_1\$ that
correspond to one and the same
eigenvalue \$r_j>0\$ {\it have one and
the same probability} in
\$\ket{\Psi}_{12}.\$ Let us denote by
\$p(Q_1^j)\$ the probability of the, in
general, composite event that is
mathematically represented by the
eigen-projector \$Q_1^j\$ of \$\rho_1\$
corresponding to \$r_j\$ (cf (9a)), and
let the multiplicity of \$r_j\$ (the
dimension of \$\cR(Q_1^j)\$) be
\$d_j.\$ Then the probability of
\$\ket{\phi}_1\bra{\phi}_1\$ is
\$p(Q_1^j)/d_j.\$ To see this, one
takes a basis
\$\{\ket{\phi_k}_1:k=1,2,\dots ,d_j\}\$
spanning \$\cR(Q_1^j),\$  or,
equivalently, \$Q_1^j=\sum_{k=1}^{d_j}
\ket{\phi_k}_1 \bra{\phi_k}_1,\$ with,
e. g., \$\ket{\phi_{k=1}}_1\equiv
\ket{\phi}_1.\$ Further, one makes use
of the {\it additivity rule of
probability}: probability of the sum of
mutually exclusive (orthogonal) events
(projectors) equals the same sum of the
probabilities of the event terms in it.

Actually, the \$\sigma$-additivity rule
of probability is the THIRD
STIPULATION. It requires that the
probability of every finite or infinite
sum of exclusive events be equal to the
same sum of the probabilities of the
event terms. We could not proceed
without it (cf subsections V.E and
V.F). The need for infinite sums will
appear four passages below.

In the {\it special case}, when
\$\rho_1\$ has only one positive
eigenvalue of multitude \$d\$ (the
dimension of the range of \$\rho_1\$),
the probability of \$\ket{\phi}_1\$ is
\$p(Q_1)/d$ (where \$Q_1\$ is the range
projector of \$\rho_1.\$ ) To proceed,
we need to evaluate \$p(Q_1)$.

To this purpose, we make the FOURTH
STIPULATION: Every state vector
\$\ket{\phi}_1\$ that belongs to the
{\it null space} of \$\rho_1\$ (or,
equivalently, when \$\ket{\phi}_1
\bra{\phi}_1\$, acting on
\$\ket{\Psi}_{12},\$ gives zero) has
{\it probability zero}. (The twin
unitaries do not influence each other
in the respective null spaces, cf
(9a,b). Hence, this assumption is
independent of the second stipulation.)

Justification for the fourth
stipulation lies in Zurek's original
framework. Namely, if the opposite
subsystem is the environment, which
establishes measurement-like
entanglement, then the Schmidt states,
e. g., the above eigen-sub-basis,
obtain partners in a Schmidt
decomposition (cf (3a)), and this leads
to measurement. States from the null
space do not appear in this, and cannot
give a positive measurement result.

One has \$1_1=Q_1+\sum_l\ket{l}_1
\bra{l}_1,\$ where \$\{\ket{l}_1:
\forall l\}\$ is a basis spanning the
null space of \$\rho_1,\$ which may be
infinite dimensional. Then,
\$p(Q_1)=p(1_1)=1\$ follows from the
third postulate (\$\sigma$-additivity)
and the fourth one. Finally, in the
above special case of only one positive
eigenvalue of \$\rho_1,\$ the
probability of
\$\ket{\phi}_1\in\cR(\rho_1)\$ is
\$1/d,$ which equals the only
eigenvalue of \$\rho_1\$ in this case.

Our next aim is to derive
\$p(Q_1^j)\$ in a more general case.\\

{\bf Stage two.} In this stage we
confine ourselves to composite state
vectors \$\ket{\Psi}_{12}\$ (i) that
have finite entanglement, i. e., the
first-subsystem reduced density
operator of which has a
finite-dimensional range; (ii) such
that each eigenvalue \$r_j\$ of
\$\rho_1\$ is a rational number.

We rewrite the eigenvalues with an
equal denominator: \$\forall j:\enskip
r_j=m_j/M.\$ Since \$\sum_jd_jr_j=1,\$
one has \$\sum_jd_jm_j=M\$ (\$d_j\$ is
the degeneracy or multiplicity of
\$r_j\$).

Now we assume that \$\ket{\Psi}_{12}\$
has a special structure:

(i) The opposite subsystem \$2\$ is
bipartite in turn, hence we replace the
notation \$2\$ by \$(2+3),\$ and
\$\ket{\Psi}_{12}\$ by
\$\ket{\Phi}_{123}.\$

(ii) a) We introduce a two-indices
eigen-sub-basis of \$\rho_1\$ spanning
the closed range \$\bar\cR(\rho_1):\$
\$\{\ket{j,k_j}_1:k_j=1,2,\dots
,d_j;\forall j\}\$ so that the
sub-basis is, as one says, adapted to
the spectral decomposition \$\rho_1=
\sum_jr_jQ_1^j\$ of the reduced density
operator, i. e., \$\forall j:\enskip
Q_1^j=\sum_{k_j=1}^{d_j}\ket{j,k_j}_1
\bra{j,k_j}_1.\$

b) We assume that \$\cH_2\$ is at least
\$M\$ dimensional, and we introduce a
basis \$\{\ket{j,k_j,l_j}_2:
l_j=1,2,\dots ,m_j; k_j=1,2,\dots
,d_j;\forall j\}\$ spanning a subspace
of \$\cH_2.\$

c) We assume that also \$\cH_3\$ is at
least \$M\$ dimensional, and we
introduce a basis
\$\{\ket{j,k_j,l_j}_3: l_j=1,2,\dots
,m_j; k_j=1,2,\dots ,d_j;\forall j\}\$
spanning a subspace of \$\cH_3.\$

d) Finally, we define via a canonical
Schmidt decomposition \$1+(2+3)$ (cf
(3b) and (5a)):
$$\ket{\Phi}_{123}\equiv \sum_j
\sum_{k_j=1}^{d_j}(m_j/M)^{1/2} \Big[
\ket{j,k_j}_1\otimes$$
$$\Big(\sum_{l_j=1}^{m_j}
(1/m_j)^{1/2}\ket{j,k_j,l_j}_2
\ket{j,k_j,l_j}_3\Big)\Big].\eqno{(13a)}$$
Equivalently, $$\ket{\Phi}_{123}\equiv
\sum_j\sum_{k_j=1}^{d_j}\sum_{l_j=1}^{m_j}
(1/M)^{1/2}\ket{j,k_j}_1\ket{j,k_j,l_j}_2
\ket{j,k_j,l_j}_3.\eqno{(13b)}$$

Viewing (13b) as a state vector of a
bipartite \$(1+2)+3\$ system, we see
that it is a canonical Schmidt
decomposition (cf (3b)). Having in mind
(5a), and utilizing the final
conclusion of stage one, we can state
that the probability of each state
vector
\$\ket{j,k_j}_1\ket{j,k_j,l_j}_2\$ is
\$1/M.\$

On the other hand, we can view (13a) as
a state vector of the bipartite system
\$1+(2+3)\$ in the form of a canonical
Schmidt decomposition. One can see that
\$\forall j,\$ \$(Q_1^j\otimes 1_2)\$
and
\$\sum_{k_j=1}^{d_j}\sum_{l_j=1}^{m_j}
\ket{j,k_j}_1\bra{j,k_j}_1\otimes
\ket{j,k_j,l_j}_2\bra{j,k_j,l_j}_2\$
act equally on \$\ket{\Phi}_{123}.\$ On
the other hand, it is easily seen that
the former projector can be written as
a sum of the latter sum of projectors
and of an orthogonal projector that
acts as zero on \$\ket{\Phi}_{123},\$
and therefore has zero probability on
account of stipulation four. Thus,
\$(Q_1^j\otimes 1_2)\$ and the above
sum have equal probabilities, which is
$$p(Q_1^j\otimes
1_2)=d_jm_j/M.\eqno{(14)}$$

As it was concluded in Stage one, the
probability of any state vector
\$\ket{\phi}_1\$ in \$\cR(Q_1^j)\$ is
\$p(Q_1^j)/d_j.\$ The projectors
\$Q_1^j\$ and \$(Q_1^j\otimes 1_2)\$
stand for the same event (viewed
locally and more globally
respectively), hence they have the same
probability in \$\ket{\Phi}_{123}.\$
Thus, \$p(\ket{\phi}_1
\bra{\phi}_1)=m_j/M=r_j,\$ i. e., it
equals the corresponding eigenvalue of
\$\rho_1$.

We see that also the eigenvalues, not
just the eigen-subspaces, i. e., the
entire operator \$\rho_1\$ is relevant
for the local probability. At this
stage we do not yet know if we are
still lacking some entity or entities.
We'll write \$X\$ for the possible
unknown.

How do we justify replacing
\$\ket{\Psi}_{12}\$ by
\$\ket{\Phi}_{123}?\$ In the state
space \$(\cH_2\otimes\cH_3)\$ there is
a pair of \ON sub-bases of
\$d=\sum_jd_j\$ vectors that appear in
(13a) (cf (15)). Evidently, there
exists a unitary operator \$U_{23}\$
that maps the Schmidt-state partners
\$\ket{j,k_j}_2\$ of \$\ket{j,k_j}_1\$
in \$\ket{\Psi}_{12}\$ tensorically
multiplied with an initial state
\$\ket{\phi_0}_3\$ into the vectors:
$$\forall k_j,\enskip\forall j:\quad
U_{23}:\quad
\ket{j,k_j}_2\ket{\phi_0}_3\enskip
\longrightarrow$$
$$\sum_{l_j=1}^{m_j}
(1/m_j)^{1/2}\ket{j,k_j,l_j}_2
\ket{j,k_j,l_j}_3.\eqno{(15)}$$ On
account of the second stipulation, any
such \$U_{23},\$ which transforms by
interaction an ancilla (subsystem
\$3\$) in state \$\ket{\phi_0}_3\$ and
subsystem \$2\$ as it is in
\$\ket{\Psi}_{12}\$ into the
\$(2+3)$-subsystem state as it is
\$\ket{\Phi}_{123}$, does not change
any local property of subsystem \$1.\$
Hence, it does not change the
probabilities either.\\

{\bf Stage three.} We make the FIFTH
STIPULATION: the sought for probability
rule is {\it continuous} in \$\rho_1,\$
i. e., if \$\rho_1=
\lim_{n\rightarrow\infty}\rho_1^n,\$
then
\$p(E_1,\rho_1,X)=\lim_{n\rightarrow
\infty}p(E_1,\rho_1^n,X),\$ for every
event (projector) \$E_1.\$  (We assume
that \$X,\$ if it exists, does not
change in the convergence process.)

Let \$\rho_1=\sum_{j=1}^Jr_jQ_1^j,\$
\$J\$ a natural number, be the spectral
form of an arbitrary density operator
with finite-dimensional range. One can
write
\$\rho_1=\lim_{n\rightarrow\infty}
\rho_1^n,\$ where
\$\rho_1^n=\sum_{j=1}^J r_j^nQ_1^j,\$
with
\$r_j=\lim_{n\rightarrow\infty}r_j^n,
\enskip j=1,2.\dots ,J,\$ and all
\$r_j^n\$ are rational numbers. (Note
that the eigen-projectors are assumed
to be the same all over the
convergence.) Then the required
continuity gives for an eigen-vector
\$\ket{r_{j_0}}\$ of \$\rho_1\$
corresponding to the eigenvalue
\$r_{j_0}\$:
\$p(\ket{r_{j_0}},\rho_1,X)=
\lim_{n\rightarrow\infty}p(\ket{r_{j_0}},
\rho_1^n,X)=r_{j_0}.\$  This extends
the conclusion of stage two to {\it all
\$\rho_1\$ with finite-dimensional
ranges}, and their eigen-vectors.

Let
\$\rho_1=\sum_{j=1}^{\infty}r_jQ_1^j\$
have an infinite-dimensional range. We
define \$\rho_1^n\equiv
\sum_{j=1}^n\Big(r_j/(\sum_{k=1}^nr_k)
\Big)Q_1^j.\$ (Note that we are taking
the same eigen-projectors \$Q_1^j.\$)
Then \$\rho_1=\lim_
{n\rightarrow\infty}\rho_1^n,\$ and for
any eigen-vector \$\ket{r_{j_0}}\$ one
has \$p(\ket{r_{j_0}},\rho_1,X)=\lim_
{n\rightarrow\infty}p(\ket{r_{j_0}},
\rho_1^n,X)=\lim_ {n\rightarrow\infty}
r_{j_0}/(\sum_{k=1}^nr_k)=r_{j_0}.\$
This extends the conclusion of the
preceding stage to {\it all reduced
density operators and their
eigen-vectors}.

As a final remark about stage three, we
point out that the continuity
postulated is meant with respect to the
so-called strong operator topology in
Hilbert space \cite{RS}. Thus, if
\$\rho =\lim_{n
\rightarrow\infty}\rho_n,\$ then, and
only then, for every vector
\$\ket{\psi}\$ one has
\$\rho\ket{\psi}=
\lim_{n\rightarrow\infty}\rho_n\ket{\psi}
.\$ This means, as well known, that
\$\lim_{n\rightarrow\infty}||\rho
\ket{\psi} -\rho_n\ket{\psi}||=0\$
(where the "distance" in the Hilbert
space is
made use of).\\

{\bf Stage four.} The result of the
preceding stages can be put as follows:
If
\$\rho_1\ket{\phi}_1=r\ket{\phi}_1,\$
then the probability is
$$p(\ket{\phi}_1,\rho_1)=r=\bra{\phi}_1
\rho_1\ket{\phi}_1.\eqno{(16)}$$ (We
have dropped \$X\$ because we already
know that, as far as eigen-vectors of
\$\rho_1\$ are concerned, nothing is
missing.) Now we wonder what about
state vectors in \$\cH_1\$ that are not
eigen-vectors of \$\rho_1$?

We make the SIXTH STIPULATION: Instead
of \$\rho_1,\$ of which the given state
\$\ket{\phi}_1\$ is not an eigen-state,
we take a different density operator
\$\rho_1'\$ of which \$\ket{\phi}_1\$
{\it is an eigenvector}, i. e., for
which
\$\rho_1'\ket{\phi}_1=r'\ket{\phi}_1\$
is valid, and which {\it is closest to
\$\rho_1\$ as such}. We stipulate that
the sought for probability is \$r'.\$
(We expect that \$r'\$ will be
determined by the requirement of
"closest as such".)

The idea behind the stipulation is the
fact that there exists non-demolition
(or repeatable) measurement, in which
the value (of the measured observable)
that has been obtained is possessed by
the system after the measurement, so
that an immediate repetition of the
same measurement necessarily gives the
same result (it is not demolished; it
can be repeated). There even exists
so-called ideal measurement in which,
if the system had a sharp value of the
measured observable before the
measurement, then it is not only this
value, but the whole state that is not
changed in the measurement. But in
general, the state (the density
operator) has to change, though
minimally, in ideal measurement. The
point is that in this change
\$\rho\enskip\rightarrow\enskip \rho'\$
the probability does not change
\$\bra{\phi}\rho'\ket{\phi}=
\bra{\phi}\rho\ket{\phi}$.

To make the requirement of "closest"
more specific, we make use of a notion
of "distance" in the set of density
operators (acting in \$\cH_1\$). As
known, the set of all linear
Hilbert-Schmidt operators in a complex
Hilbert space is, in turn, a complex
Hilbert space itself (cf Appendix C).
All density operators are
Hilbert-Schmidt operators. Every
Hilbert space is a distantial space,
and "closest" is well defined in it.

We are not going to solve the problem
of finding the closest density operator
to \$\rho_1\$ because a related problem
has been solved in previous work of the
author \cite{AnnPhys69}. Namely, the
fact that \$\ket{\phi}_1\$ is an
eigenvector of \$\rho_1'\$ can be put
in the equivalent form of a mixture
$$\rho_1'=r'\ket{\phi}_1\bra{\phi}_1
+$$ $$(1-r')\Big[\Big(\ket{\phi}_1
\bra{\phi}_1\Big)^{\perp}\rho_1'
\Big(\ket{\phi}_1
\bra{\phi}_1\Big)^{\perp}\Big/(1-r')
\Big].\eqno{(17)}$$ In (17) \$\rho_1'\$
is a mixture of two states, one in
which \$\ket{\phi}_1 \bra{\phi}_1\$
{\it as an observable} has the sharp
value \$1,\$ and one in which it has
the sharp value \$0\$.

In Ref. \cite{AnnPhys69} it was shown
that when a density operator \$\rho_1\$
is given, the closest density operator
\$\rho_1',\$ among those that satisfy
(17), is:
$$\rho_1'\equiv \bra{\phi}_1\rho_1\ket{
\phi}_1\ket{\phi}_1\bra{\phi}_1+$$ $$
\Big(\ket{\phi}_1\bra{\phi}_1\Big)^{\perp}
\rho_1
\Big(\ket{\phi}_1\bra{\phi}_1\Big)^{\perp}.
\eqno{(18)}$$ Thus, $$r'=
\bra{\phi}_1\rho_1\ket{ \phi}_1,
\eqno{(19)}$$ and the same formula (the
last expression in (16)) extends also
to the case when \$\ket{\phi}_1\$ is
not an eigenvector of \$\rho_1$.

Incidentally, the requirement of
closest \$\rho'\$ to \$\rho\$ under the
restriction that the "closest" is taken
among those density operators that are
mixtures of states with sharp values of
the measured observable
\$A=\sum_ka_kP_k\$ (spectral form)
defines the L\"{u}ders state \$\rho'=
\sum_kP_k\rho P_k\$ \cite{AnnPhys69}.
(It was postulated \cite{Lud}; and as
such it appears in textbooks
\cite{Messiah}.) As well known, in
ideal measurement \$\rho\$ changes to
the L\"{u}ders state. (In so-called
selective ideal measurement, when one
takes the subensemble corresponding to
a specific result, say, \$a_{k_0},\$
the change of state is \$\rho\enskip
\rightarrow\enskip P_{k_0}\rho P_{k_0}
\Big/\tr(P_{k_0}\rho ).\$ This is
sometimes called "the projection
postulate".)

As a final remark on stage four, one
should point out that "distance" in the
Hilbert space of linear Hilbert-Schmidt
operators also defines a topology, in
particular a convergence of density
operators. It is stronger than the
so-called strong operator topology
utilized in the preceding stage. More
about this in Appendix C.\\

{\bf Stage five.} Finally, we have to
find out what should be the probability
rule when \$\rho \$ is not an improper,
but a proper mixture, i. e., when there
are no correlations with another
system. We take first an isolated pure
state \$\ket{\psi}.\$

We start with an infinite sequence of
correlated bipartite state vectors
\$\{\ket{\Psi_{12}}^n:n=1,2,\dots
,\infty \}\$ such that, as far as the
reduced density operator is concerned,
one has
$$\forall n:\quad
\rho_1^n=(1-1/n)\ket{\psi}_1\bra{\psi}_1+
$$ $$
\Big(\ket{\psi}_1\bra{\psi}_1\Big)^{\perp}
\rho_1^n\Big(\ket{\psi}_1\bra{\psi}_1\Big)
^{\perp},\eqno{(20)}$$ where
\$\ket{\psi}_1\$ actually equals
\$\ket{\psi}.\$ (It is well known that
for every density operator \$\rho_1\$
there exists a state vector
\$\ket{\Psi}_{12}\$ such that
\$\rho_1=\tr_2\Big(\ket{\Psi}_{12}
\bra{\Psi}_{12}\Big).\$ This claim is
easily proved using the spectral form
(5a) of \$\rho_1\$ and the canonical
Schmidt decomposition (3b).) We now
write index \$1\$ because we now do
have correlations with subsystem \$2.\$

Obviously
$$\ket{\psi}_1\bra{\psi}_1=
\lim_{n\rightarrow
\infty}\rho_1^n.\eqno{(21)}$$

According to our fifth stipulation, the
probability rule is continuous in the
density operator. Hence, $$\forall
\ket{\phi}:\quad p\Big(\ket{\phi},
\ket{\psi}\Big)=\lim_{n\rightarrow
\infty}p\Big(\ket{\phi}_1,
\rho_1^n\Big)=$$ $$\lim_{n\rightarrow
\infty}\bra{\phi}_1\rho_1^n\ket{\phi}_1=
\bra{\phi}_1\lim_{n\rightarrow \infty}
\rho_1^n\ket{\phi}_1.$$ This finally
gives $$\forall \ket{\phi}:\quad
p\Big(\ket{\phi},\ket{\psi}\Big)=
\bra{\phi}\Big(\ket{\psi}\bra{\psi}
\Big)\ket{\phi}=
|\bra{\phi}\ket{\psi}|^2.\eqno{(22)}$$
In this way, the same probability rule
is extended to isolated pure states.

If \$\rho\$ is an isolated mixed state,
i. e., a proper mixture, one can take
any of its (infinitely many)
decompositions into pure states, say,
$$\rho=\sum_kw_k\ket{\psi_k}\bra{\psi_k},
$$ where \$w_k\$ are the statistical
weights (\$\forall k:\enskip
w_k>0;\enskip \sum_kw_k=1\$). Then
$$p\Big(\ket{\phi},\rho\Big)=
\sum_kw_k\bra{\phi}\Big(\ket{\psi_k}
\bra{\psi_k}\Big)\ket{\phi}.$$ This
finally gives
$$p\Big(\ket{\phi},\rho\Big)=
\bra{\phi}\rho\ket{\phi},\eqno{(23)}$$
extending the same probability rule to
mixed isolated states. (It is obvious
that the choice of the above
decomposition into pure states is
immaterial. One can take the spectral
decomposition e. g.)\\

\section{RELATION TO THE LITERATURE}

This article comes after 8 studies of
thought-provoking analiticity
\cite{Zurek1}, \cite{Zurek2},
\cite{Zurek3}, \cite{Zurek4},
\cite{Schlossh2}, \cite{Barnum2},
\cite{Mohrhoff}, \cite{Caves} on
Zurek's derivation of Born's rule. It
has profited from most of them.

The purpose of this section is not to
review these articles; the purpose is
to contrast some ideas from 5 of these
works with the present version in order
to shed more light on the latter.

\subsection{SCHLOSSHAUER-FINE}

For the purpose of a logical order in
my comments, I'll mess up the order of
the quotations from the article of \Sc
on Zurek's argument \cite{Schlossh2}.

\Sc are inspired to define the precise
framework for Zurek's endeavor and try
to justify it saying (DISCUSSION, (A)):

\begin{quote}
"Apart from the problem of how to do
cosmology, we might take a pragmatic
point of view here by stating that any
observation of the events to which we
wish to assign probabilities will
always require a measurement-like
context that involves an open system
interacting with an external observer,
and that therefore the inability of
Zurek's approach to derive
probabilities for a closed, undivided
system should not be considered as a
shortcoming of the argument."
\end{quote}

This may well be the case. In the
present version, one views the
probability rule as a potential
property of the system. Measurement is
something separate; it comes afterwards
when an observer wants to get
cognizance of the probabilities. The
present study is an attempt to view
Zurek's argument in such a setting of
ideas. Incidentally, in the present
version one can no longer speak of an
"inability of Zurek's approach to
derive probabilities for a closed,
undivided system".

Besides, the "problem of how to do
cosmology" is considered by many
foundationally minded physicists to be
an important problem in modern
quantum-mechanical thinking. After all,
interaction with the environment and
decoherence that sets in (a phenomenon
to which Zurek gave an enormous
contribution) is primarily
observer-independent (though it may
contain an observer), and it fits well
into quantum cosmology. The present
study envisages Zurek's argument in a
measurement-independent
and observer-independent way.\\

In their CONCLUDING REMARKS \Sc say:

\begin{quote}
"...a fundamental statement about any
probabilistic theory: We cannot derive
probabilities from a theory that does
not already contain some probabilistic
concept; at some stage, we need to "put
probabilities in to get probabilities
out".
\end{quote}

In the present version of the theory, a
realization of this pessimistic
statement can be seen in the assumption
that local probabilities exist at all
(in the first stipulation, (b)), and in
the application of additivity (and
\$\sigma$-additivity) of probability
(the third stipulation). Incidentally,
the quoted claim of \Sc is perhaps only
mildly pessimistic \cite{FN2Mohrhoff}\\

As a counterpart of the stipulations in
the present version, \Sc state (near
the end of their INTRODUCTION):

\begin{quote}
"...we find that Zurek's derivation is
based at least on the following
assumptions:

(1) The probability for a particular
outcome, i. e., for the occurrence of a
specific value of a measured physical
quantity, is identified with the
probability for the eigenstate of the
measured observable with eigenvalue
corresponding to the measured value -
an assumption that would follow from
the {\it eigenvalue-eigenstate link}.

(2) Probabilities of a system \$\cS\$
entangled with another system \$\cE\$
are a function of the {\it local}
properties of \$\cS\$ only, which are
exclusively determined by the state
vector of the {\it composite} system
\$\cS\cE$.

(3) For a composite state in the
Schmidt form \$\ket{\psi_{\cS\cE}}=
\sum_k\lambda_k\ket{s_k}\ket{e_k},\$
the probability for \$\ket{s_k}\$ is
{\it equal} to the probability for
\$\ket{e_k}$.

(4) Probabilities associated with a
system \$\cS\$ entangled with another
system \$\cE\$ remain {\it unchanged}
when certain transformations (namely,
Zurek's "envariant transformations")
are applied that only act on \$\cE\$
(and similarly for \$\cS\$ and \$\cE\$
interchanged)."
\end{quote}

Assumption (1) is very important. It is
the quantum logical approach. (See the
comment on it in section V.B
.)
Assumption (2) is reproduced in the
present version as the first
stipulation.

Having in mind the above quotation on
"putting in and taking out
probability", assumption (3) was
carefully avoided in the present
version, which goes beyond the Schmidt
decomposition. In the approaches that
hang on to the decomposition, and all
preceding ones are such, putting in
probability where it is equal to \$1\$
seems unavoidable.

As to assumption (4), it is, to my
mind, {\it the basic idea} of Zurek's
argument. Though \Sc "consider Zurek's
approach promising" (INTRODUCTION),
they feel very unhappy about this basic
assumption (DISCUSSION, F2):

\begin{quote}
"...we do not see why shifting features
of \$\cE\$ , that is, doing something
to the environment, should not alter
the "guess"... an observer of \$\cS\$
would make concerning \$\cS$-outcomes.
\end{quote}

\Sc point to Zurek's desire to bolster
his argument by a subjective aspect
with an observer who observes only
subsystem \$\cS,\$ but who is aware of
the composite state vector
\$\ket{\Psi}_{\cS\cE}.\$ This observer
"makes guesses" and "attributes
likelihood" to state vectors
\$\ket{\phi}_{\cS}.\$ \Sc make critical
comments on this aspect.

Weighing if the subjective aspect at
issue is useful or even justified is
avoided in the present version. It was
assumed that Zurek's argument can do
without it (cf the comment on Caves's
first-quoted remark about this).

\Sc finish the quoted passage saying:

\begin{quote}
"Here, if possible, one would like to
see some further argument (or
motivation) for why the probabilities
of one system should be immune to swaps
among the basis states of the other
system."
\end{quote}

Apparently, locality or
subsystem-property is a basic
stipulation (the first stipulation in
the present version), i. e., the basic
idea how Zurek envisages probability.
Naturally, one may object that it is
hindsight, because we know the
probability rule, and it implies the
locality idea.

When thinking of quantum ideas without
the probability rule, as Zurek does,
why not try to insert into them a local
probability idea? The motivation lies
in our intuitive expectation to find
nature with as many local properties as
possible (to enable us to do physics).
After all, the well known tremendous
reaction of the scientific community to
Bell's theorem dealing with subquantum
locality is an impressive indication of
how important locality is considered to
be.

Envariance, or twin unitaries in the
present equivalent formulation, (and
broader, see the second stipulation)
provide us with a means to {\it define}
what it means "local" or a "subsystem
property" when the reduced density
operator is devoid of physical meaning
to begin with, and we do not know what
the state of the subsystem is. The two
subsystems \$\cS\$ and \$\cE\$ are {\it
remote} from each other. This means
that they cannot dynamically influence
each other. To put it in more detail,
no ancilla (or measuring instrument)
interacting with subsystem \$\cE\$ can
have any dynamical influence on the
opposite subsystem \$\cS$.

Now, isn't it natural to stipulate with
Zurek, that subsystem or local
properties of \$\cS\$ are those
properties that cannot be changed by
"doing something" to the opposite
subsystem (action of an ancilla
included), or otherwise the property
would be global? (It might be useful to
point out that the essential role of
locality in Zurek's derivation is made
clear also in his "facts" (cf the sixth
quotation in subsection IV.C),
especially in fact 2.)\\

As to the parenthetical final remark of
\Sc in assumption (4) (of the third
quotation), the present version did not
make use of "interchanged" roles of
\$\cS\$ and \$\cE.\$ Entanglement
"treats" the two subsystems in a
symmetrical way. So the interchange is
quite all right, but it
was felt, in expounding the present
version,  that it was unnecessary.\\

\Sc say (DISCUSSION, (G)):

\begin{quote}
"According to Zurek, ...the observer is
aware of the "menu" of possible
outcomes..."
\end{quote}

In the present version, one is after
 a local probability rule and, to
start with, one has no other idea what
"local" means, except what envariance
gives. Gradually, one endows the
reduced density operator of the
subsystem with the known standard
physical meaning. It seems that this
gradual building up knowledge of what
"local" means for probabilities is in
Zurek's wording handled by the
imaginary observer to whom, besides
\$\ket{\Psi}_{\cS\cE},\$ only the
subsystem \$\cS\$ is accessible. But
what is the "subsystem"? The state
space \$\cH_{\cS}\$ and the state
vectors in it are all that is at the
imaginary observer's disposal and at
ours to start to build the "subsystem"
notion. This is Zurek's "menu" (in the
understanding of the present author).

Perhaps, one should stress that, if one
envisages probability as a
potentiality, as it is done in the
present approach, then it seems natural
to take in the "menu" {\it all} state
vectors \$\ket{\phi}_{\cS};\$ not just
those that are eigen-vectors of the
reduced density operator
\$\rho_{\cS},\$ which, at the
beginning, has almost no physical
meaning. ("Almost" is inserted in view
of the Second Theorem on twin
unitaries.) Contrariwise, if one
envisages probabilities in the process
of measurement (or observation), as
Zurek does (and his commentators follow
him), then taking the Schmidt
decomposition is the suitable
procedure. In the present version, this
is avoided (except in the mathematical
interlude, in deriving the properties
of twin unitaries in subsection II.A).
\\

In the last passage of the DISCUSSION
of \Sc the basis of the opposite
subsystem that appears in the Schmidt
decomposition is subjected to
though-provoking critical comments.
This is one of the reasons why the
present version kept clear of the
Schmidt decomposition.\\

As to the eigenvalue-eigenstate link
given in assumption (1) (third
quotation), \Sc say (DISCUSSION, (C)):

\begin{quote}
"Clearly, from the point of view of
observations and measurements, we would
like to assign probabilities to the
occurrence of the specific values of
the observable \$\cO\$ that has been
measured, i. e.,to the "outcomes". The
eigenvalue-eigenstate link of \QM
postulates that a system has a value
for an observable \IF the state of the
system is an eigenstate characteristic
of that value (or a proper mixture of
those eigenstates)."
\end{quote}

In the preceding section it was assumed
that events are represented by
projectors. This is {\it the quantum
logical approach} (because projectors
can be interpreted as events,
properties or logical statements), in
which the projectors are more
elementary than observables.
(Mathematically, one constructs
Hermitian operators out of projectors
using the spectral theorem.)
Physically, the yes-no experiments
carry the essence of \qm. The quantum
logical approach is resumed in
subsection V.B(a). (Zurek, in his Phys.
Rev. paper, seems to be trying to take
a more general approach: he is dealing
with potential future records.)

On the other hand, observables and
their eigenvalues ("outcomes") are the
standard or textbook starting point for
probabilities. Utilizing the
eigenvalue-eigenstate link, leading to
the quantum logical standpoint, is a
choice of approach, which has to be
justified in the end. Namely, when the
probability rule is finally available,
the eigenvalue-eigenstate link is a
theorem: A state (density operator)
\$\rho\$ has the sharp value \$o\$ of
an observable \$\cO\$ \IF (i) the
former is an eigenvalue of the latter
and (ii) \$\rho,\$ when written as any
mixture (possibly a trivial one)e
states, it consists only of eigen-
states of \$\cO\$ corresponding to this
eigenvalue (cf the Introduction in
\cite{Specific}).

Finally, it should be pointed out what
has been taken over from the article
\cite{Schlossh2} of \sc. The second
quotation led to caution concerning
"putting in" as little probability as
possible. It was the reason for
avoiding the use of the Schmidt
decomposition and hence also assumption
3 (in the third quotation). The last
quotation gave rise to thoughts about
the non-contextuality involved (cf
subsection V.B).

\subsection{Barnum}

In what follows a few comments  in
connection with Barnum's reaction
\cite{Barnum2} to Zurek's derivation of
probability will be given.

Barnum says (p.2, left column):

\begin{quote}
"In our opinion, the version of Zurek's
argument we give below does not depend
crucially on whether measurement is
interpreted in this way (relative state
interpretation, F. H.), or as involving
"collapse", or in some other way (for
example as involving "collapse" of our
knowledge, say  in a process similar to
Bayesian updating
\cite{Bayes})."
\end{quote}

Hopefully, also the version of Zurek's
argument expounded in the preceding
section is independent of the existence
or non-existence of objective
"collapse" in nature. (As to purely
subjective "Bayesian updating", it is
hard to see what one can update if
nothing happened in nature. Let us be
reminded of John Bell's famous dictum:
"Information? Whose information,
information about what?" But, some of
us may just be incorrigible realists,
"whatever realism means" - as the late
Rudolph Peierls used to say.)

Assuming the existence of objective
collapse, there are two remote effects
due to entanglement: distant
measurement \cite{DistMeas}, or more
generally, remote ensemble
decomposition \cite{FH06}, and remote
preparation \cite{Schroed2},
\cite{genSteer}, \cite{FH06} (the
selective aspect of the former). It all
started with Schr\"{o}dinger
\cite{Schroed2}, who pointed out that
doing a suitable selective measurement
on subsystem \$2,\$ one can "steer"
(his word for remote preparation) the
remote system \$1\$ into any state
\$\ket{\phi}_1\$ that is an element of
the range of \$\rho_1,\$ but with a
certain positive probability.
(Schr\"{o}dinger assumed that the range
was finite dimensional. This was
extended to
\$\ket{\phi}_1\in\cR(\rho_1^{1/2})\$ in
\cite{genSteer} for infinite
dimensional ranges, and the maximal
probability, i. e., the best way to do
remote preparation, was evaluated
recently \cite{FH06}.)

Neither Schr\"{o}dinger
\cite{Schroed1}, \cite{Schroed2}, nor
anyone in the Belgrade group who worked
on his program of "disentanglement"
\cite{DistMeas}, \cite{saY},
\cite{FH06} has ever, to the best of
the present author's knowledge, tried
to utilize remote preparation for an
argument of probability because this
would be "putting probability in to get
probability out" (cf the second
quotation in the preceding subsection),
i. e., an evidently circular argument.

It is a beauty of Zurek's argument that
envariance, or remote unitary operation
if one takes twin unitaries (the other
face of envariance), has no probability
at the start. It is deterministic: You
perform a \$U_2\$ local transformation
on the opposite subsystem, and {\it
ipso facto} one gets deterministically
the transformation \$U_1\$ on the
subsystem that is investigated. So,
Zurek seems to be quite right that this
concept can be used to shed light on
the quantum probability notion (as far
as it is assumed to be local).\\

One gets the impression that Barnum
feels that his insistence on {\it no
signalling} and {\it symmetric roles}
that \$\cS\$ and \$\cE\$ should play is
an important improvement on Zurek's
argument. In particular, Barnum says
(p. 2, right column):

\begin{quote}
"Perhaps, however, there is a stronger
argument for no \$\cS$-to-$\cE\$
signalling in relative state
interpretation. On such an
interpretation, once macroscopic
aspects of \$\cE\$  have been
correlated with \$\cS\$ (the system has
been "measured" by an observer who is
part of \$\cE\$), the ability to affect
probabilities of components of the
state in subspaces corresponding to
those distinct macroscopic aspects of
\$\cE,\$ by manipulating \$\cS,\$
jeopardizes the interpretation of these
numbers as "probabilities" at all. ...
(within a generally subjectivist
approach to probability in its aspect
as something to be {\it used} in
science and everyday life..., an
approach to which I am rather
partial),..."
\end{quote}

Barnum is, of course, consistent. The
purpose of quoting this passage is
mostly to underline the difference in
the approaches to Zurek's argument by
Barnum and the present version. Namely,
in the latter an attempt is made to
keep the remote influence in one
direction only, as Zurek originally
did. Not because Barnum appears to be
wrong; it is because the one-direction
approach is considered simpler. There
is another difference: Barnum says to
be partial to subjectivism, and the
present author has confessed above to
be a realist. (This is not in the sense
to negate or underestimate
subjectivism. But the latter is
understood by the present author as
subjective cognizance of objective
reality.)\\

Barnum says (p. 3, both columns):

\begin{quote}
"...if the joint state \$\cS\cE\$ is
viewed as the outcome of a measurement
"in the Schmidt basis" on \$\cS,\$ by
an environment \$\cE\$ that includes
the observer, whose "definite
measurement results" line up with the
Schmidt basis for \$\cE,\$ ascribing
probabilities to these suffices for
ascribing probabilities to "definite
measurement results" ..."
\end{quote}

Also \Sc pointed to this feature of
Zurek's argument of "putting in
probability" in \$\cE\$, and "getting
out" probability in \$\cS\$ (cf the
second quotation and assumption 3 in
the third quotation in the preceding
subsection). Apparently, Zurek "puts
in" no more than (probabilistic)
certainty. This certainly is not
circularity. Nevertheless, the present
version takes another route.\\

There is another aspect of the present
version that it shares with Zurek's
original one. It is assuming
non-contextuality. But let us first see
what Barnum says on the subsject (p. 3,
right column):

\begin{quote}
"Note that we have not yet established
that, for a given state, the
probabilities of components in
subspaces are {\it independent} of the
subspace decomposition in which they
occur, an assumption similar to that
made in Gleason's theorem, and which
might allow us to use Gleason's theorem
as part of an argument for quantum
probabilities. Of course, a potential
virtue of the argument from envariance
is precisely that it does not make any
such assumption to begin with."
\end{quote}

One is here on quantum-logical grounds.
{\it Quantum-logical non-contextuality}
means, in the understanding of the
present author, that if \$F\$ is a
composite event (the projector project
onto a more-than-2 dimensional
subspace), then no matter in which of
the infinitely many possible ways \$F\$
is written as a sum of mutually
exclusive (orthogonal) elementary
events (ray projectors), and defined in
this way, the probability of \$F\$ is
one and the same. This is so on account
of \$\sigma$-additivity. (See also the
discussion in subsection V.B(a)).

It is hard to see how one can avoid the
quantum-logical non-contextuality in
Zurek's argument. Namely, when one
wants to evaluate the probabilities of
the equally probable states
\$\ket{\phi}_1\$ that correspond to one
and the same eigenvalue of \$\rho_1\$
(stage one in the preceding section),
one cannot avoid using additivity.
Besides, also in the evaluation of the
probability of the eigen-event \$Q_1\$
(the range projector) when \$\rho_1\$
has only one positive eigenvalue
requires the use of additivity (and the
zero-probability assumption, cf the
third and the fourth stipulations in
the preceding section). Then, as it was
argued in the preceding passage,
quantum-logical non-contextuality has
been utilized. (More on this in
subsections V.B and V.E. See also
subsection V.F.)

Gleason gives the complete answer (cf
subsection V.F). Then what is the point
of Zurek's argument? I'll attempt an
answer to this worrisome question in
the concluding comments in
the next section (see subsection V.F).\\

After the quoted passage, Barnum writes
about, what he calls, the Perfect
Correlation Principle. From the point
of view of the Belgrade group, he talks
about twin observables (cf subsection B
on twin Hermitians in section II.): The
measurement of any subsystem observable
that is compatible (commuting) with the
corresponding reduced density operator
is {\it ipso facto} also a measurement
(so-called distant measurement) of a
twin observable on the opposite
subsystem.\\

Barnum further says, speaking of Stan
and Emma instead of subsystems, and
applying his \$\cS\rightarrow\cE\$
no-remote-influence ("no signalling")
approach (p. 3, right column):

\begin{quote}
"Whether or not Stan measures anything
should be immaterial to Emma's
probability, by no-signalling."
\end{quote}

Twin Hermitians are mathematically very
closely connected with twin unitaries
(subsection B in section II.). Distant
measurement can make non-contextuality
very plausible for suitable, i. e.,
with the reduced density operator
compatible, subsystem observables. But
distant measurement is derived from the
probability rule in \qm. This way one
cannot avoid circularity.

Subsystem observables {\it not
compatible} with the corresponding
density operator do not give rise to
distant measurement; they cause distant
ensemble decomposition (see
\cite{FH06}). Here we are outside
envariance, i. e., we are using
subsystem unitaries (in the sense of
subsection II.B) that do not have a
twin.\\

On his page 5, left column, Barnum
discusses at length Zurek's assumption
of continuity of probability as a
function of \$\rho_{\cS}\$ . Among
other things, he says:

\begin{quote}
"It is not clear to us why one would
rule out discontinuous probability
assignments even though they may seem
"pathological"."
\end{quote}

In the preceding section "continuity"
entered as the fifth stipulation. It
has led, in the end, to the quantum
probability rule. The argument
presented leaves open the possibility
that also probability that is not
continuous in \$\rho\$ might exist. But
we know from Gleason's theorem that,
though he does assume continuity in the
projectors (via \$\sigma$-additivity as
a strengthening of additivity, cf
subsection V.E), he does not assume
continuity in \$\rho.\$ Thus,
probability discontinuous in \$\rho\$
does not seem to exist.\\

The present author is especially
indebted to Barnum for his useful
suggestion about how to extend Zurek's
argument to state vectors
\$\ket{\phi}_1\$ that are not
eigenvectors of \$\rho_1.\$ He
suggested (in private communication):
"Perhaps one could get somewhere by
making assumptions about probabilities
zero and one..." This fitted in well
with the theorem from previous work on
the closest suitable state, i. e.,
state of zero and one probabilities (cf
the sixth stipulation in section III of
this article and relation (17)).

Finally, it should be stated what is
the main insight gained from the
article \cite{Barnum2} of Barnum. It
confirmed the suspicion, stemming from
Zurek's writings, that the concrete
idea of system and environment can be
generalized to any entangled
subsystems. (Stan and Emma achieve
this.) The continuity assumption is not
as trivial as one might think. Barnum
made me give a lot of thought to the
quantum-logical non-contextuality (cf
subsection V.B(a)), and the relation
between Gleason's theorem
and Zurek's argument (cf subsection V.F).\\

\subsection{Zurek's most mature
article on envariance}

Zurek in his most mature, Physical
Review, article \cite{Zurek4} takes
into account the comments of \Sc and
Barnum. The exposition of the preceding
section will now be put in relation to
Zurek's original argument presented
there. (Quotations will be taken from
pages in the archive
copy, version 2.)\\

In the abstract Zurek says:

\begin{quote}
"Probabilities derived in this manner
(he means from envariance, F. H.) are
an objective reflection of the
underlying state of the system - they
represent experimentally verifiable
symmetries, and not just a subjective
"state of knowledge" of the observer."
\end{quote}

In the present version, one confines
oneself to this attitude of the founder
of envariance, though he finishes the
abstract as follows.

\begin{quote}
"Envariant origin of Born's rule for
probabilities sheds a new light on the
relation between ignorance (and hence
information) and the nature of quantum
states."
\end{quote}

On p. 1, left column he completes this
thought as follows:

\begin{quote}
"The nature of "missing information"
and the origin of probabilities in
quantum physics are two related themes,
closely tied to its interpretation."
\end{quote}

One cannot but fully agree with this.
The subjective side of Zurek's argument
has, nevertheless, been disregarded in
the present version because
considerably more than the basic
quantum formalism has been made use of
in it (unlike in the preceding
versions), and, hence, it is quite
intricate as it is.\\

On p. 1, left column, Zurek says:

\begin{quote}
"We shall, however, refrain from using
"trace" and "reduced density matrix".
Their physical significance is based on
Born's rule....,to avoid
circularity,..."
\end{quote}

In contrast to Zurek's original
version, in the present one not only
that "trace" and "reduced density
matrix" are not avoided, they are the
mathematical starting point.
Admittedly, they are at the start
physically devoid of meaning. But the
second theorem on twin unitaries (the
other face of envariance) in subsection
A of section II. discloses the
relevance of these concepts for
envariance. Since one of the basic
ideas of Zurek is that the
probabilities in the system \$\cS\$ are
{\it local}, and we do not have the
reduced density matrix \$\rho_{\cS}\$
determining the subsystem state and
thus defining locality, it appears
natural to use envariance (twin
unitaries) for the definition of what
is local. Then, the {\it mathematical}
notion of the reduced density matrix
turns out to be relevant, and
gradually, taking the steps of Zurek's
argument, the reduced density matrix
becomes endowed with the standard
physical meaning.\\

At the beginning of his argument, on p.
2, right column, Zurek lines up the
basic assumptions of "bare" \QM (or \QM
without collapse): that the universe
consists of systems, each of which has
a state space; that the state space of
composite systems are tensor products;
and that the unitary dynamical law is
valid. (See also Zurek's three spelled
out "Facts" - the sixth quotation
below.) All these were
tacitly assumed in section III.\\

At the beginning of the left column, p.
3, Zurek says:

\begin{quote}
"We shall call the part of the global
state that can be acted upon to affect
such a restoration of the preexisting
global state {\it the environment
\$\cE\$}. Hence, the {\it
environment-assisted invariance}, or -
for brevity - envariance. We shall soon
see that there may be more than one
such subsystem. In that case we shall
use \$\cE\$ to designate their union."
\end{quote}

It appears that Zurek envisages,
actually, more-or-less the whole
universe , or at least, a large part of
it containing all systems that have
ever interacted with the subsystem
\$\cS\$ at issue. In contrast to this,
the version of the argument in section
III laid emphasis on the existence of
entanglement with any opposite
subsystem (but cf subsection V.D). Any
larger system \$(1+2)\$ in any
entangled state \$\ket{\Psi}_{12}\$
that has one and the same local or
first-subsystem probability would do.
Since subsystem \$2\$ is arbitrary, it
can also be the
environment as Zurek envisages it.\\

On p. 4, left column, Zurek lists three
"facts", which he considers basic to
his approach.

\begin{quote}
"{\bf Fact 1:} Unitary transformations
must act on the system to alter its
state. (That is, when the evolution
operator does not operate on the
Hilbert space \$\cH_{\cS}\$ of the
system, i. e., when it has a form
\$\dots\otimes {\bf 1}_{\cS}\otimes
\dots\$ the state of \$\cS\$ remains
the same.)

{\bf Fact 2:} The state of the system
\$\cS\$ is all that is needed (and all
that is available) to predict
measurement outcomes, including their
probabilities.

{\bf Fact 3:} The state of a larger
composite system that includes \$\cS\$
as a subsystem is all that is needed
(and all that is available) to
determine the state of the system
\$\cS$."
\end{quote}

Zurek adds "... the above {\bf facts}
are interpretation-neutral and the
states (e. g., 'the state of \$\cS\$')
they refer to need not be pure."

I find Zurek's "facts" fully
acceptable, and I have tacitly built
them into the present approach (like
the above basic assumptions of the
no-collapse part of \qm). Actually, his
broad "state" concept helped me to
decide to stick to the reduced density
operator \$\rho_1,\$ the physical
relevance of which is suggested by the
two theorems on twin unitaries in
subsection II.A. As it could be seen in
section III, Zurek's argument enables
one to endow the mathematical concept
of the reduced density operator
gradually with the standard physical
meaning yielding the quantum
probability
rule.\\

On p. 4, left column, Zurek says:

\begin{quote}
"Indeed, Schmidt expansion is
occasionally defined by absorbing
phases in the states which means that
all the non-zero coefficients end up
real and positive ... . This is a
dangerous oversimplification. Phases
matter... ."
\end{quote}

Zurek is, of course, quite clear about
the role of canonical Schmidt
decomposition (see section II.A above).
What he means, I believe, is that one
must be careful about phases in any
expansion of the global state; one can
disregard them only after a careful
analysis as the one he presents. Since
the present version goes beyond the
Schmidt decomposition, it turned out
that the separate question of phases
actually does not come up.\\

On the other hand, one can fully accept
his words (p. 4, bottom of right
column):

\begin{quote}
"Lemma 3 we have just established is
the cornerstone of our approach."
\end{quote}

His Lemma 3 is about envariant swaps of
orthogonal first-subsystem eigenstates
of \$\rho_1,\$ and, later in his
Theorem 2., it implies their equal
probability. In methodological contrast
to Zurek's Lemma 3, in section III
above the second theorem on twin
unitaries (section II.A) was used to
establish equal probability of any two
state vectors in one and the same
eigensubspace of \$\rho_1.\$ But, this
is, of course, equivalent to Zurek's
Theorem 2.\\

On p. 5, left column, Zurek gives a
very nice discussion of the
complementarity between knowledge of
the whole and knowledge of the part -
{\it complementarity of global and
local due to entanglement}
. There was
no need to
enter this in the present version.\\

On p. 7, right column, Zurek says:

\begin{quote}
"Let us also assume that states that do
not appear in the above superposition
(i. e., appear with Schmidt coefficient
zero) have zero probability. (We shall
motivate this rather natural assumption
later in the paper.)"
\end{quote}

This is the fourth stipulation in
section III. This is "rather natural"
when we already know the quantum rule
of probability. In Zurek's setting of
no such knowledge, it appears to come
out of the blue. But a stipulation can
do this.

Zurek resumes this question on p. 19,
left column, considering a rather
intricate composite state "representing
both the fine-grained and the
coarse-grained records". He essentially
describes observation or measurement in
my understanding. He says:

\begin{quote}
"The form of ... (the composite state,
F. H.) justifies assigning zero
probability to ... (state vectors of
the system, F. H.) that do not appear,
- i. e., appear with zero amplitude -
in the initial state of the system.
Quite simply, there is no state of the
observer  with a record of such
zero-amplitude Schmidt states of the
system ... (in the composite state, F.
H.)."
\end{quote}

This is convincing in the context of
Zurek's objective probabilities - as he
calls them. If probability is treated
as a potentiality, no matter if it will
be ever measured or not, as it is in
the present approach, then one had
better not use this argument. (It is
used only as a plausibility
justification in the present version.)\\

On p. 7, right column, Zurek says:

\begin{quote}
"Moreover, probability of any subset of
\$n\$ mutually exclusive events is
additive. ... We shall motivate also
this (very natural) assumption of the
additivity of probabilities further in
discussion of quantum measurements in
Section V (thus going beyond the
starting point of e. g. Gleason ...)"
\end{quote}

Zurek has stated (on p. 5, left column)
that he will use, besides envariance,
also "a variety of small subsets of
natural assumptions". At this place of
his text, it appears that additivity of
probability is one of them. Actually,
it is a very strong assumption on the
quantum-logical ground(cf the
discussion of this in subsections
V.B(a) and V.E). One can accept that
the measurement context makes it more
plausible, but it still is an extra
assumption.

Zurek resumes this question on pp. 18
and 19. He is at pains to derive
"additivity of probability from
envariance". He says:

\begin{quote}
"To demonstrate Lemma 5 (a key step in
his endeavor, F. H.) we need one more
property - the fact that when a certain
event \$\cU\$ \$(p(\cU)=1)\$ can be
decomposed into two mutually exclusive
events, \$\cU=k\vee k^{\perp},\$ their
probability must add up to unity:
$$p(\cU)=p(k\vee k^{\perp})=p(k)+
p(k^{\perp})=1.$$ This assumption
introduces (in a very limited setting)
additivity. It is equivalent to the
statement that "something will
certainly happen"."
\end{quote}

We have discussed above the \Sc comment
"you put in probability, to get out
probability". Zurek's just quoted
passage looks somewhat similar: you put
in additivity, to get out additivity
(though you put it in "in a very
limited setting", but at the crucial
place). This question is resumed
in detail in subsection V.E.\\

Zurek starts his subsection D. of
section II. stating that he will
"complete derivation of Born's rule" by
considering the case of unequal
absolute values of the coefficients in
the Schmidt decomposition. Clearly,
unlike section III of this paper, Zurek
had no intention to go further than
encompassing the eigenvectors of
\$\rho_1.$ In his terminology, that is
"Born's rule".\\

Zurek finishes section II., after he
has discussed rational moduli of
Schmidt coefficient (which has been
completely taken over in section III
above) saying:

\begin{quote}
"This is Born's rule. The extension to
the case where \$|a_k|^2\$ (the moduli,
F. H.) are incommensurate is
straightforward by continuity as
rational numbers are dense among
reals."
\end{quote}

This seems to be another of Zurek's
"natural assumptions". In the present
version, it was raised to the level of
a stipulation following the convincing
discussion of Barnum (cf the last
quotation and the last passage in the
preceding subsection).\\

Zurek's section V is devoted to a
rederivation of Born's rule from
envariance. In his section II. the
environment \$\cE\$ could and needed
not contain an observer. He didn't
actually make use of him. In section V
the observer is explicitly made use of
(consistent with, e. g., the
relative-state theory of Everett
\cite{Everett}). One gets the feeling
that this exposition, in which it is
explicit that Zurek is after
probability in the process of
measurement (or observation), is more
convincing and successful.

In the present version, measurement is
"off limits" (as Zurek would say). Twin
unitaries (the other face of
envariance) are a direct consequence of
entanglement (cf subsection II.A of
this article). In the present version,
Zurek's argument was treated as strong
enough to carry out the complete
program: quantum probability rule from
entanglement, treating the former as a
potentiality. This standpoint is,
apparently, in keeping with the
following passage of Zurek's paper.

On p. 23, left column, Zurek says:

\begin{quote}
"...even when one can deduce
probabilities {\it a priori} using
envariance, they better be consistent
with the relative frequencies estimated
by the observer {\it a posteriori} in
sufficiently large samples. ... We
shall conclude that when probabilities
can be deduced directly from the pure
state (he means \$\ket{\Psi}_{\cS\cE},
\$ F. H.), the two approaches are in
agreement , but that the {\it a priori}
probabilities obtained from
envariance-based arguments are more
fundamental."
\end{quote}

Precisely so! Because probabilities are
an {\it a priori} notion, and "more
fundamental" than the relative
frequencies, in terms of which they are
measured, the probabilities should be
treated as a {\it potentiality}.\\

Finally, it is needless to state what
has been learn't from Zurek. The entire
theory is his. The rest of us are only
conjuring up different variations on it
to gain a deeper grasp of the matter.\\

\subsection{Mohrhoff}

I'll begin with the abstract of
Mohrhoff's paper \cite{Mohrhoff} on
Zurek's "Born's rule from envariance"
argument, which lacks Zurek's Physical
Review paper (discussed in the
preceding subsection), and both
Barnum's article and the one of Caves
in its references. Mohrhoff says:

\begin{quote}
"Zurek claims to have derived Born's
rule noncircularly... from
deterministically evolving quantum
states. ... this claim is exaggerated
if not wholly unjustified. ...it is not
sufficient to assume that quantum
states are somehow associated with
probabilities and then prove that these
probabilities are given by Born's
rule."
\end{quote}

Mohrhoff calls in question the, as he
puts it, "so-called derivation" of
Born's rule. Strictly logically,
"derivation" of a claim means that the
claim is {\it a necessity}. Now,
probabilities are a necessity in a
deterministically evolving universe
from a physical point of view as made
clear in section V of Zurek's Phys.
Rev. paper. But logically, Mohrhoff is
right that one assumes the existence of
probabilities, and then one finds out
what they look like. The present
version is certainly not better than
that.\\

Mohrhoff even strengthens his critical
attitude on p. 4 (the archive version
is taken) after having shortly reviewed
Zurek's argument:

\begin{quote}
"What is thereby proved is that {\it
if} quantum states are associated with
probabilities then Born's rule holds.
But how do quantum states come to be
associated with probabilities? As long
as this question remains unanswered,
one has not elucidated the origin of
probabilities in quantum physics, as
Zurek claims to have done."
\end{quote}

In spite of Zurek's wording in
expounding his argument, he does not
appear to be claiming to have answered
Mohrhoff's "question"; the present
version certainly has not. One becomes
pessimistic at this point, and one is
inclined to partially agree with
Mohrhoff's first sentence in his
Introduction:

\begin{quote}
"In any metaphysical framework that
treats quantum states as
deterministically evolving ontological
states, such as Everett's many-worlds
interpretation, Born's rule has to be
postulated."
\end{quote}

Zurek's derivation of Born's rule
suggests that this claim should be
weakened be replacing "Born's rule" in
it by "probability".\\

In the following quotation (bottom of
p. 6), Mohrhoff hits at the very
foundation of Zurek's argument.

\begin{quote}
"The rather mystical-sounding statement
that knowledge about the whole implies
ignorance of the parts (he means
complementarity of global and local, F.
H.) is thus largely a statement about
correlated probability distributions
over measurement outcomes. Given its
implicit reference to probabilities, it
does not elucidate the "origin of
probabilities" but rather shows that
probabilities are present from the
start, however cleverly they may be
concealed by mystical language."
\end{quote}

As far as correlated probability
distributions are concerned, Mohrhoff
has a point. Indeed, the remote
effects, which can be, in principle,
either immediately confirmed by
coincidence measurement or subsequently
by a suitable measurement on the
opposite (remote) subsystem, are
observationally nothing else than {\it
correlated probabilities}.

Does this ruin Zurek's argument? I
think not at all. Complementarity of
global and local is a well known fact.
Besides, {\it entanglement} should be
understood as another peculiar {\it
potentiality}, which can lead to the
potentiality of probability. After all,
the latter is what Zurek is after (at
least as it is understood in the
present version). Hopefully, these
potentialities are not just "mystical
language" "concealing" the true state
of affairs (cf subsection
V.C).\\

Mohrhoff's rejection of Zurek's
argument is rather deep-rooted. On p. 7
he says:

\begin{quote}
"To my mind, the conclusion to be drawn
from the past failures (including
Zurek's) to derive probabilities
noncircularly from deterministically
evolving ontological quantum states, is
that quantum states are probability
measures and should not be construed as
evolving ontological states. Theorists
ought to think of them the way
experimentalists use them, namely, as
algorithms for computing the
probabilities of possible measurement
outcomes  on the basis of actual
measurement outcomes."
\end{quote}

It seems that Mohrhoff has accepted
Bohr's standpoint that ontology in
quantum physics is metaphysics, i. e.,
beyond physics, perhaps philosophy.
Mohrhoff has even strengthened Bohr's
rejection of a nowadays rather widely
accepted ontology speaking of
"pseudophysics" (or false physics). He
seems to be, what one sometimes calls,
an "instrumentalist" believing only in
the reality of the laboratory
instruments; the rest is "mystical

language" \cite{FNMohrhoff}. This calls
to mind Mermin's, perhaps somewhat
unjust, nickname for such a standpoint:
"the shut up and calculate
interpretation of \qm" (cf the article
by \sc).

Though Mohrhoff stands at the farthest
from the ontological standpoint of
Zurek and the rest of his commentators
(including the present author), his
criticism and objections should be
taken seriously. After all, ontology is
also a potentiality; if one does not
believe in it, you
can't prove it.\\

Finally, let it be stated what has been
learnt from Mohrhoff's article. His
scepticism about the non-circularity of
Zurek's argument (cf the first
quotation, and especially the second
one) helped to decide to try to treat
probability as a potentiality (without
any measurement or observation). Next,
following Mohrhoff's explicit warning
(see his third quotation), the present
version postulates the existence of
probability (as part of the first
postulate). Mohrhoff's uncompromising
attitude is a challenge that has led to
an attempt to put Zurek's argument in a
transparently non-circular way. To what
extent the present version has
succeeded in this will be discussed
again in the next
section (cf subsection V.C).\\

\subsection{Caves}

Caves' reaction \cite{Caves} to Zurek's
argument appeared with all the
references that have been commented
upon so far.

At the very beginning of his treatise,
Caves reacts to the Phys. Rev. Letters
version, and comments on Zurek's
subjective standpoint saying:

\begin{quote}
"It is hard to tell from WHZ's
(Zurek's, F. H.) discussion whether he
sees his derivation as justifying the
Born rule as the way for an observer to
assign subjective probabilities or as
the rule for objective probabilities
that adhere within a relative state."
\end{quote}

Later on, Caves quotes the same as in
my first quotation in the subsection on
Zurek's Phys. Rev. paper, and decides
that "WHZ is thinking in terms of
objective probabilities". In the
present version the subjective side of
the problem is completely omitted, but
it should be emphasized that this is
not because it is not considered
important.

Though sometimes it is hard to see
one's way through Zurek's "underbrush
of verbiage" (as Caves says for Barnum)
in his copious expositions (the
exposition in the present article is
probably no better), it is clear that
Zurek's approach to fundamental
problems is rather all-encompassing. In
particular, he, no doubt, recognizes
that no thorough ontology can disregard
epistemology. But in the latter, the
observer's cognition is a reflection of
reality. When an observer cannot
distinguish two envariantly swapable
states, e. g., this means, that they
are objectively indiscernible, i. e.,
equal, etc. (I am sure, Caves sees the
work of Zurek in a similar manner, but
he seems to object to the way how Zurek
unfolds his ideas.)\\

On p. 2, Caves starts with a simple
(non-composite) system \$A,\$ and a
non-trivial observable for it. He then
points out that Zurek considers the
unitary evolution corresponding to
interaction with an ideally measuring
apparatus \$B.\$ (Ideal measurement is
not only a non-demolition one, i. e.,
result preserving, but also eigen-state
preserving, and, of course, probability
preserving.) This fits well into the
sixth stipulation of the present
version, in which the closest suitable
state is the L\"{u}ders state
corresponding precisely to ideal
measurement.\\

Caves further says on p. 2:

\begin{quote}
"Notice that what I am saying is that
in WHZ's approach, it is the Schmidt
relative state that {\it defines} the
notion of outcomes for system \$A;\$
without the entanglement with system
\$B\$, one cannot even talk about
outcomes for the basis
\$\{\ket{a_k}\}\$ (the eigenbasis of
the measured observable, F. H.)."
\end{quote}

Zurek "derives" probabilities from
entanglement, and the latter he
displays in terms of a Schmidt
decomposition. No re-definition of
events takes place here. (One can read
in Zurek's Phys. Rev. article a
detailed discussion on how events,
pointer states, etc. emerge from
correlations.)

Caves further says (on the same page):

\begin{quote}
"... it has already been assumed that
the probabilities that he is seeking
... have no dependence on the
environmental states \$\ket{b_k}\$
(partners of \$\ket{a_k}\$ in the
Schmidt decomposition, F. H.). This is
a kind of foundational noncontextuality
assumption that underlies the whole
approach. I will call it {\it
environmental noncontextuality} for
lack of a better name."
\end{quote}

This is an attempt to view Zurek's
derivation from another angle. In
section III of this article a rather
different, though essentially
equivalent view was presented. Perhaps,
one should be reminded of it. The
probabilities in subsystem \$A\$ (to
use Caves' notation for the first
subsystem), though defined by the
bipartite entangled state
\$\ket{\psi}_{AB},\$ are actually {\it
locally} determined. Then the rest of
the argument goes on in utilizing twin
unitaries (the other face of
envariance) to find this local
determination. Naturally, by the very
fact of local determination of
subsystem probability (the first
stipulation), the details of the
opposite subsystem (the environment)
don't really matter. Therefore, no
emphasis was put on Cave's
"environmental non-contextuality".\\

On p. 3 Caves says:
\begin{quote}
"WHZ wants to view envariance as the
key to his derivation, but it is just a
way to write the consequences of
environmental non-contextuality, when
they provide any useful constraints, in
terms of system unitaries, instead of
environment unitaries. It turns out not
to be necessary to translate
environmental non-contextuality to
system unitaries for any of the steps
in the derivation."
\end{quote}

The last statement seems to be the most
important one in Caves' article; it
appears to be the program of his
version of Zurek's argument. And he
carries it out in the rest of his
paper.\\

In Caves' version, as in all the other
versions, Schmidt decomposition is
adhered to as the only widely known way
how to handle entanglement. As a
consequence, it turns out indispensable
to put some probability in the
environment, to get out probability in
the system. It is assumption (3) in the
article of \sc; Barnum calls it the
Perfect Correlation Principle (same as
"twin observables" in the work of the
Belgrade group); Zurek uses it and
emphasizes that probability-one
statements are put in; Caves accepts
Barnum's term. It consists simply in
equal probabilities of the partners in
a Schmidt decomposition. Both Barnum
and Caves make use of the environment
in a way that is more than necessary
from the point of view of the present
approach. Namely, on p. 4 Caves says:

\begin{quote}
"The point is that WHZ's derivation
depends on an unstated assumption that
one can interchange the roles of
systems \$A\$ and \$B\$ in the case of
Schmidt states with amplitudes of equal
magnitude."\\
\end{quote}

In contrast to the rest of the authors
of versions commented upon so far,
Caves couldn't readily accept the
suitable extension of the environment
to reduce unequal Schmidt coefficients
to equal ones. On p. 6 he says:

\begin{quote}
"We were originally told that the very
notion of outcomes for system \$A\$
required us to think about a joint pure
state with the appropriate Schmidt
decomposition. Now we are told that the
notion of outcomes requires us to think
about a much more complicated
three-system joint state, where the two
additional systems must have a
dimension big enough to accommodate the
rational approximation to the desired
probabilities. Does this mean the
notion of outcomes depends on the value
of the amplitudes? This is a very
unattractive alternative, so what we
really must think is that for all
amplitudes, the notion of outcomes
requires us to think in terms of  a big
three-system joint state, where \$B\$
and \$C\$ have arbitrarily large
dimensions. We are now supposed to
believe that the notion of outcomes for
system \$A\$ requires us to think in
terms of two other systems correlated
in a particular way, which has no
apparent relation to the number of
outcomes of system \$A.\$ Even a
relative-state believer would find this
hard to swallow, and it makes the
Perfect Correlations Principle
assumption far less natural, because
this construction wrecks the
nice-looking symmetry between \$A\$ and
the systems to which it is coupled and
even between \$AB\$ and \$C.\$ It is a
heck of a lot less attractive than the
original picture we were presented and
really should have been stated at the
outset."
\end{quote}

This rebellious passage of Caves was of
great help in realizing that one should
not confine oneself to unitaries of the
opposite system that have a twin for
the system under consideration treating
locality. Also broader
opposite-subsystem unitaries cannot
change what is local in the system (see
the second stipulation in section III
of this article), and hence are part of
the definition of the subsystem state
and local properties. Then interaction
with a suitable ancilla, which takes
place in terms of such a unitary, comes
natural, and subsystem \$A\$ of the
enlarged system \$A+BC\$ that Caves is
objecting to still has the same
locality or subsystem state, and
the same subsystem probabilities.\\

Caves closes his consideration on p. 6
saying:

\begin{quote}
"In the end one is left wondering what
makes the envariance argument any more
compelling than just asserting that a
swap symmetry means that a state with
equal amplitudes has equal
probabilities and then moving on to the
argument that extends to rational
amplitudes."
\end{quote}

One should bear in mind that the swap
symmetry is equivalent to symmetry
under the group of twin unitaries,
which is, in turn, equivalent to the
essence of the envariance argument.

Finally, it should be pointed out that
the need for broader opposite-subsystem
unitaries than just those \$U_2\$ that
have a twin \$U_1\$ (see the second
stipulation in the present version) is
not the only thing that has been learnt
from Caves' article \cite{Caves}. His
comments raised the question how to
extend Zurek's argument to isolated
systems. (A solution using continuity
is presented in the present approach.)

\section{CONCLUDING REMARKS}

There are some points that require
additional clarification and comment.

\subsection{Summing up the
stipulations of the present version}

The FIRST STIPULATION is: (a) Though
the given pure state
\$\ket{\Psi}_{12}\$ determines all
properties in the composite system,
therefore also all those  of subsystem
\$1,\$ the latter must be {\it
determined actually by the subsystem
alone}. This is, by (vague) definition,
what is meant by {\it local}
properties.

{\it (b)} There exist local or
subsystem probabilities of all
elementary events
\$\ket{\phi}_1\bra{\phi}_1,\$
\$\ket{\phi}_1\in\cH_1$.

The SECOND STIPULATION is that
subsystem or {\it local properties must
not be changeable by remote action}, i.
e., by applying a second-subsystem
unitary \$U_2\$ to \$\ket{\Psi}_{12}\$
or any unitary \$U_{23}\$ applied to
the opposite subsystem with an ancilla
(subsystem \$3\$).

The most important part of the precise
mathematical formulation of the second
stipulation is in terms of twin
unitaries (cf (8a)). No local unitary
\$U_1\$ that has a twin \$U_2\$ must be
able to change any local property.

The \$\sigma$-additivity rule of
probability is the THIRD STIPULATION.
It requires that the probability of
every finite or infinite sum of
exclusive events be equal to the same
sum of the probabilities of the event
terms.

The FOURTH STIPULATION: Every state
vector \$\ket{\phi}_1\$ that belongs to
the {\it null space} of \$\rho_1\$ (or,
equivalently, when \$\ket{\phi}_1
\bra{\phi}_1\$ acting on
\$\ket{\Psi}_{12},\$ gives zero) has
{\it probability zero}. (The twin
unitaries do not influence each other
in the respective null spaces, cf
(9a,b). Hence, this assumption is
independent of the second stipulation.)

The FIFTH STIPULATION: the sought for
probability rule is {\it continuous} in
\$\rho_1,\$ i. e., if \$\rho_1=
\lim_{n\rightarrow\infty}\rho_1^n,\$
then
\$p(E_1,\rho_1,X)=\lim_{n\rightarrow
\infty}p(E_1,\rho_1^n,X),\$ for every
event (projector) \$E_1,\$ and \$X\$
stands for the possible yet unknown
additional entity needed for a complete
local probability rule. Further we
assume that \$X,\$ if it exists, does
not change in the convergence process.

The SIXTH STIPULATION: Instead of
\$\rho_1,\$ of which the given state
\$\ket{\phi}_1\$ is not an eigen-state,
we take a different density operator
\$\rho_1'\$ of which \$\ket{\phi}_1\$
{\it is an eigenvector}, i. e., for
which
\$\rho_1'\ket{\phi}_1=r'\ket{\phi}_1\$
is valid, and which {\it is closest to}
\$\rho_1\$ as such. We stipulate that
the sought for probability is \$r'$.\\

Comparing the stipulations to Zurek's
facts (sixth quotation in subsection
IV.C), we see that facts 3 and 2
strictly correspond to the first
stipulation (a). (Fact 1 is connected
with answering the question in
subsection V.G.)\\

Let us compare the 6 stipulations with
the 4 assumptions  of \Sc (cf the third
quotation from their article).
Assumption (1) is not among the former,
because I understand Zurek's starting
point is quantum logical, and so is
mine. Zurek does not seem to consider
observables, and neither am I.

Assumption (3) is avoided because of
the possible suspicion that it is
"putting probability in" (cf the second
quotation from \sc) though Zurek
remarks that it is no more than putting
probability-one statements in.

Three assumptions that, apparently,
cannot be avoided, have been raised to
the status of stipulations: that of
\$\sigma$-additivity, that of null
probability of the null-space vectors
\$\ket{\phi}_1,\$ and, finally that of
continuity. (The sixth stipulation in
the present version is, of course, not
covered by \Sc because they did not
consider extending Zurek's argument.)\\

\subsection{Non-contextuality in the
quantum logical approach}

{\it (a) The event non-contextuality.}
From the quantum logical point of view,
the elementary events occur in only one
way. There is no question of context.
But on account of the implication
relation in the structure of all events
(the projector \$E\$ implies the
projector \$F,\$ i. e., \$E\leq F\$ \IF
\$EF=E\$) every composite event can
occur as a consequence of the
occurrence of different elementary
events that imply it. Nevertheless, the
probability does not depend on this.

As a matter of fact, the probabilities
of the composite events are in  Section
III of this article, following Zurek,
defined in terms of mutually exclusive
elementary events (orthogonal
ray-projectors, each defined by a state
vector) using \$\sigma$-additivity.

{\it (b) Non-contextuality with respect
to observables.} A given elementary (or
composite) event can, in general, be
the eigen-event (eigen-projector) of
different observables. (This,
essentially, amounts to the so-called
eigenvalue-eigen-state link.)
Correspondingly, the event can occur in
measurement of different observables.
The probability of the
event does not depend on this.\\

\subsection{Circularity?}

In the second quotation from the
article of \sc, the curse of a
"fundamental statement" that one cannot
"get probability out" of a theory
unless one "puts some probability in"
should be valid also for the present
version. It appears to be valid no more
for the present version of Zurek's
argument than for Gleason's theorem.
Namely, what both "put in" is the
assumption that probability exists and
that \$\sigma$-additivity is valid for
it.

Let us return to Mohrhoff's attempt of
a fatal blow at Zurek's argument in the
last but one quotation from his article
stating that entanglement itself is
correlation of probabilities. Hence,
using entanglement as a starting point
means "putting probability in". No
wonder that one "gets probability out".

One can hardly shatter Mohrhoff's
criticism. It all depends on how much
belief one is prepared to put in
theory. Taking an extremely
positivistic attitude, one can say
that, e. g., "interference" is all that
exists in the phenomenon when one sees
it; "coherence" in the quantum
mechanical formalism giving rise to
interference is, according to such a
point of view, just a part of the
formalism without immediate physical
meaning.

If one decides, however, to allow some
reality to theoretical concepts, then,
in the case at issue, "entanglement" is
a theoretical concept (the correlation
operator in the present approach), a
potentiality, which is believed to be
real in nature. We can observe its
consequence as correlation of
probabilities, but it is
more than that.\\

\subsection{The role of entanglement}

In the present version, entanglement
enters through, what was said to be,
the sole entanglement entity - the
correlation operator \$U_a\$ (see the
correlated subsystem picture in section
II.A.). In terms of this entity the
first theorem on twin unitaries (near
the end of section II.A.) gives a
complete answer to the question which
unitaries have a twin, and which
opposite-subsystem unitary is the
(unique) twin.

In section III, in unfolding the
present version, the correlation
operator (and hence entanglement) was
not made use of at all. All that was
utilized was the general form of a
first-subsystem unitary that has a
twin: \$U_1=\sum_jU_1^jQ_1^j+U_1Q_1
^{\perp},\$ where \$1_1=\sum_jQ_1^j+
Q_1^{\perp}\$ is the eigen-resolution
of the unity with respect to (distinct
eigenvalues) of the reduced density
operator \$\rho_1 \Big(\equiv\tr_2(
\ket{\Psi}_{12}\bra{\Psi}_{12})\Big),\$
and \$\forall j:\enskip U_1^j\$ is an
arbitrary unitary in the eigen-subspace
\$\cR(Q_1^j)\$ corresponding to the
positive eigenvalue \$r_j\$ of
\$\rho_1\$ (cf (9a)). (In the necessity
part of the proof, \$U_a\$ was not
used; it was used only in the
sufficiency part.)

These unitaries (Zurek's envariance
unitaries) are utilized to establish
what are local or first-subsystem
properties, in particular, local
probabilities. It immediately follows
that any two distinct eigen-vectors
corresponding to the same eigenvalue of
\$\rho_1\$ determine equal-probability
events (cf Stage one in section III).
Thus, envariance is made use of in the
first and most important step of
Zurek's argument in a completely
assumption-of-probability-free way.

Nevertheless, twin unitaries
(envariance) is due to entanglement,
and Zureks argument is based on the
latter. Entanglement is, as well known,
the basic staff of which quantum
communication and quantum computation
are made of. No wonder that
entanglement is increasingly considered
to be a fundamental physical entity. As
an illustration for this, one may
mention that preservation of
entanglement has been proposed as an
equivalent second law of thermodynamics
for composite systems (cf Ref.
\cite{Popescu} and the references
therein).

\subsection{\$\sigma$-additivity}

To get an idea how "heavy" the
\$\sigma$-additivity assumption for
probability intuitively is, we put it
in the form of a "staircase" of
gradually strengthened partial
assumptions.

The starting point is the fact is that
if any event \$F\$ occurs, the opposite
event \$F^{\perp}\$ \$\Big(\equiv
(1-F)\Big)\$ does not occur (in
suitable measurement, of course).

1) It is plausible to assume that
\$F+F^{\perp}=1\$ has
\$p(F)+p(F^{\perp})=1\$ as its
consequence in any quantum state.

2) If \$E+F=G\$ (all being events, i.
e., projectors, and \$EF=0\$), then, in
view of the fact that, e. g., \$F\$ is
the opposite event of \$E\$ {\it in}
\$G,\$ i. e., \$F=E^{\perp}G,\$ and in
view of assumption (1), it is plausible
to assume that \$E+F=G\$ implies
\$p(E)+p(F)=p(G)\$ in any quantum
state. Obviously, assumption (2) is a
strengthening of assumption (1).

{\bf Lemma.} Assumption (2) implies
additivity for every finite orthogonal
sum of events:
\$\sum_iE_i=G\enskip\Rightarrow
\enskip\sum_ip(E_i)=p(G)\$ in any
quantum state.

{\bf Proof.} If the lemma is valid for
\$n\$ terms, then
$$p\Big(\sum_{i=1}^{(n+1)}
E_i\Big)=p\Big((\sum_{i=1}^nE_i)+E_{(n+1)}
\Big)=$$
$$p\Big(\sum_{i=1}^nE_i\Big)+p(E_{(n+1)})=
\sum_{i=1}^{(n+1)}p(E_i),$$ i. e., it
is valid also for \$(n+1)\$ terms. By
assumption, it is valid for two terms.
By total induction, it is then valid
for every finite sum.\hfill $\Box$\\

3) If
\$G=\lim_{n\rightarrow\infty}F_n\$ and
the sequence \$\{F_n:n=1,2,\dots ,
\infty\}\$ is non-descending (\$\forall
n: F_{(n+1)}\geq
F_n\enskip\Leftrightarrow\enskip
F_{(n+1)}F_n=F_n\$), then the
assumption of {\it continuity} in the
probability
\$p(G)=\lim_{n\rightarrow\infty}p(F_n)\$
is plausible (otherwise one could have
jumps in probability and no event
responsible for it). Assuming the
validity of assumption (2), it implies
$$p(\sum_{i=1}^{\infty}E_i)=
p(\lim_{n\rightarrow\infty}\sum_{i=1}^n
E_i)=$$ $$\lim_{n\rightarrow\infty}
\sum_{i=1}^np(E_i)=\sum_{i=1}^{\infty}
p(E_i),$$ i. e., \$\sigma$-additivity
ensues.

If one wants to estimate how "steep"
each of these "stairs" is, one is on
intuitive ground burdened with feeling
and arbitrariness. Assumption (1) seems
to be the largest "step" (with respect
to the stated fact that is its
premise). Once (1) is given, assumption
(2) (equivalent to additivity of
probability) seems very natural, hence
less "steep". The final assumption (3)
seems even more natural, and hence
least "steep".

At one place Zurek admits that (1) is
an assumption (cf the last-but-two
quotation in the subsection on Zurek's
article). One wonders if he can avoid
to assume (2). Leaning on "the standard
approach of Laplace" \cite{Laplace}
(second passage, right column, p. 18,
\cite{Zurek4}), in which "by
definition" "the probability of a
composite event is a ratio of the
number of favorable equiprobable events
to the total", property (2) of
probability follows. Zurek seems to
adopt this reasoning to a large extent
within eigen-subspaces \$\cR(Q_1^j)\$
of \$\rho_1\$ (cf (7c) in this
article). Thus, partially he can avoid
to assume (2). But can he do this
generally?

The form
\$\bra{\phi}_1\rho_1\ket{\phi}_1\$ of
the probability rule achieved,
following Zurek, in the present version
(shortly, the present form), is
equivalent to the (much more generally
looking) trace rule precisely on
account of \$\sigma$-additivity. Taking
an infinitely composite event \$E=
\sum_{i=1}^{\infty}\ket{i}\bra{i},\$
\$\sigma$-additivity allows to
transform the present form into the
trace rule:
$$p(E)=\sum_{i=1}^{\infty}
\bra{i}\rho\ket{i}= \sum_{i=1}^{\infty}
\tr(\rho\ket{i}\bra{i})=\tr(\rho E).$$
Thus, without \$\sigma$-additivity the
present form is not the standard
probability rule.

Besides, the argument just presented
can appear in the very context of
Zurek's argument. Let
\$\ket{\Psi}_{12}\$ be infinitely
entangled, or, equivalently, let
\$\rho_1\$ have an infinitely
dimensional range. Further, let the
above set \$\{\ket{i}_1:i=1,2,\dots
,\infty\}\$ (with index) be a set of
eigenvectors of \$\rho_1\$
(corresponding to different
eigenvalues), but let they not span the
whole range \$\bar\cR(\rho_1).\$
Without the validity of
\$\sigma$-additivity the present rule
does not give an answer what is the
probability \$p(E_1,\rho_1),\$ where
\$E_1\equiv\sum_{i=1}^{\infty}
\ket{i}_1\bra{i}_1.\$ Thus, if one want
the general form of the probability
rule, and in the present version
nothing less is wanted, then one must
assume (2) and the continuity in (3).

\subsection{Zurek's argument and
Gleason's theorem}

In an effort to tighten up Zurek's
argument, his "small natural" and some
tacit assumptions have been avoided as
much as possible. The most disquieting
consequence was raising
\$\sigma$-additivity to the status of a
stipulation. This was no different than
in Gleason's well known theorem
\cite{Gleason}, which goes as follows.

One assumes that one has a map
associating a number \$p\$ from the
doubly-closed interval \$[0,1]\$ with
every subspace, or, equivalently, with
every projector \$F\$ (projecting onto
a subspace) observing
\$\sigma$-additivity, i. e.\
$$p(\sum_iF_i)=\sum_ip(F_i)\eqno{(24a)}$$
for every orthogonal decomposition
(finite or infinite) of every
projector. Then, for every such map,
there exists a unique density operator
\$\rho\$ such that
$$p(F)=\tr(F\rho )\eqno{(24b)}$$ for
every projector (the trace rule). Thus,
the set of all density operators and
that of all quantum probabilities stand
in a natural one-to-one relation.

Logically, this makes the other five
stipulations (besides
\$\sigma$-additivity) in the present
version of Zurek's argument
unnecessary. Barnum is on to this (see
the above fourth quotation from his
article), but his understanding seems
to be that Zurek's assumption of
additivity is weaker than that of
Gleason. At least in the present
version this is not so.

Let us be reminded that in Stage one of
section III additivity had to be used
in concluding that if
\$\rho_1\ket{\phi}_1=r_j\ket{\phi}_1,\$
and the corresponding eigen-projector
is \$Q_1^j,\$ projecting onto a
\$d_j$-dimensional subspace (which is
necessarily finite), then the
probability of \$\ket{\phi}_1\$ is
\$p(Q_1^j)/d_j$.

Further, \$\sigma$-additivity  had to
be used in Stage two to conclude that
\$p(Q_1^j)=r_jd_j,\$ where also the
fourth postulate about zero
probabilities from the (possibly
infinite dimensional) null space of
\$\rho_1\$ had to be utilized. ("Had to
be" means, of course, that "the present
author saw no other way".)

Zurek's argument is very valuable
though we have the theorem of Gleason.
Perhaps a famous dictum of Wigner can
help to make this clear. When faced
with the challenge of computer
simulations to replace analytical
solutions of intricate equations of
important physical meaning, Wigner has
allegedly said "I am glad that your
computer understands the solutions; but
I also would like to understand them."

\Sc say (in the Introduction to their
paper):

\begin{quote}
"...Gleason's theorem is usually
considered as giving rather little
physical insight into the emergence of
quantum probabilities and the Born
rule."
\end{quote}

As to the logical necessity of "the
emergence of quantum probabilities", it
seems hopeless (unless if the
probabilities would prove subjective,
i. e., due to ignorance, like in
classical physics, after all). Neither
Gleason, nor Zurek, nor anybody else -
as it seems to me - can derive
objective quantum probability, in the
sense to show that it necessarily
follows from deterministic \qm. But,
once one realizes from physical
considerations that probability must
exist, then one makes the logical
assumption that it exists, and then one
wonders what its form is.

Gleason gives the complete answer at
once in the form of the trace rule. One
can then derive from it the other five
postulates of the present version and
more. To use Wigner's words, the
mathematics in the proof of Gleason's
theorem "understands" the uniqueness
and the other wonders of the quantum
probability rule, but we do not.

Now, the extra 5 stipulations in the
present version (besides
\$\sigma$-additivity), though logically
unnecessary in view of Gleason's
theorem, nevertheless, thanks to
Zurek's ingenuity, help to unfold
before our eyes the simplicity and full
generality of the quantum rule in the
form \$\bra{\phi}\rho\ket{\phi}$
(equivalent
to the trace rule).\\

\subsection{Why unitary operators?}

Both envariance and its other face,
unitary twins, are expressed in terms
of unitary operators. One can raise the
question in the title of the
subsection.

The answer lies in the notion of {\it
distant influence}. One assumes that
the nearby subsystem \$1\$ is
dynamically decoupled from another
subsystem \$2,\$ but not statistically.
Quantum correlations are assumed to
exist between the two subsystems. On
account of these correlations one can
manipulate subsystem \$2\$ in order to
make changes in subsystem \$1\$
(without interaction with it). By
definition, local are those properties
of the nearby subsystem that cannot be
changed by the described distant
influence. Probabilities of events on
subsystem \$1\$ were stipulated to be
local.

One is thinking in terms of so-called
bare \qm, i. e., \QM without collapse.
Then all conceivable manipulations of
the distant subsystem are unitary
evolutions (suitable interactions of
suitably chosen subsystems - all
without any interaction with subsystem
\$1\$ ). As Zurek puts it in his Fact 1
(sixth quotation in subsection IV.C):
"Unitary transformations must act on
the system to alter its state." (This
goes for the distant subsystem which
should exert the distant influence.)

Unitary evolution preserves the total
probability of events. The suspicion
has been voiced that the restriction to
unitary operators might just be a case
of "putting in probability in order to
get out probability" \cite{Max}. Even
if this is so, it appears to be even
milder than Zurek's "putting in"
probability-one assumptions (cf last
passage in subsection B.1 in
\cite{Zurek4}).

One may try to argue that the unitarity
of the evolution operator (of the
dynamical law) does not contain any
probability assumption. Namely, one may
start with the Schr\"{o}dinger
equation, of which the unitary
evolution operator is the integrated
form (from instantaneous tendency of
change in a finite interval). At first
glance, the Schr\"{o}dinger equation
has nothing to do with probabilities.
But this is not quite so. The dynamical
law, instantaneous or for a finite
interval, gives the change of the
quantum state, which is, in turn,
equivalent to the totality of
probability predictions.

Perhaps one should not expect to derive
probabilities exclusively from other
notions (cf the second quotation from
Ref. 2 in subsection IV.A).\\

{\bf APPENDIX A}

We prove now that the correlation
operator \$U_a\$ is independent of the
choice of the eigen-sub-basis of
\$\rho_1\$ (cf (5a)) that spans
\$\bar\cR(\rho_1)\$ in which the strong
Schmidt decomposition of
\$\ket{\Psi}_{12}\$ (cf (3c)) is
written.

Let \$\{\ket{j,k_j}_1:\forall
k_j,\forall j\}\$ and
\$\{\ket{j,l_j}_1:\forall l_j,\forall
j\}\$ be two arbitrary eigen-sub-bases
of \$\rho_1\$ spanning
\$\bar\cR(\rho_1).\$ The vectors are
written with two indices, \$j\$
denoting the eigen-subspace
\$\cR(Q_1^j)\$ to which the vector
belongs, and the other index \$k_j\$
(\$l_j\$) enumerates the vectors within
the subspace.

A proof goes as follows. Let
$$\forall j:\quad \ket{j,k_j}_1=
\sum_{l_j}U_{k_j,l_j}^{(j)}\ket{j,l_j}_1,$$
where \$\Big(U_{k_j,l_j}^{(j)}\Big)\$
are unitary sub-matrices. Then, keeping
\$U_a\$ one and the same, we can start
out with the strong Schmidt
decomposition in the
\$k_j$-eigen-sub-basis, and after a few
simple steps (utilizing the
antilinearity of \$U_a\$ and the
unitarity of the transition
sub-matrices), we end up with the
strong Schmidt decomposition (of the
same \$\ket{\Psi}_{12}\$) in the
\$l_j$-eigen-sub-basis:

$$\ket{\Psi}_{12}=\sum_j\sum_{k_j}
r_j^{1/2}\ket{j,k_j}_1 \Big(U_a
\ket{j,k_j}_1\Big)_2=$$
$$\sum_j\sum_{k_j}\Big\{r_j^{1/2}
\Big(\sum_{l_j}U_{k_j,l_j}^{(j)}
\ket{j,l_j}_1\Big)\otimes$$ $$\Big[U_a
\Big(\sum_{l_j'}
U_{k_j,l_j'}^{(j)}\ket{j,l_j'}_1\Big)
\Big]_2\Big\}=$$
$$\sum_j\sum_{l_j}\sum_{l_j'}\Big\{
r_j^{1/2}
\Big(\sum_{k_j}U_{k_j,l_j}^{(j)}U_{k_j,
l_j'}^{(j)*}\Big)\ket{j,l_j}_1\otimes$$
$$ \Big(U_a\ket{j,l_j'}_1\Big)_2\Big\}=
\sum_j\sum_{l_j}\sum_{l_j'}\Big\{r_j^{1/2}
\delta_{l_j,l_j'}\ket{j,l_j}_1\otimes$$
$$ \Big(U_a\ket{j,l_j'}_1\Big)_2\Big\}=
\sum_j\sum_{l_j}r_j^{1/2}
\ket{j,l_j}_1\Big(U_a\ket{j,l_j}_1
\Big)_2.$$\hfill $\Box$\\

{\bf APPENDIX B}

We elaborate now the {\it group of
pairs of unitary twins}.

Let \$(U_1',U_2')\$ and \$(U_1,U_2)\$
be two pairs of twin unitaries for a
given bipartite state vector
\$\ket{\Psi}_{12},\$ i. e., let
\$U_1'\ket{\Psi}_{12}=U_2'\ket{\Psi}_{12},
\$ and
\$U_1\ket{\Psi}_{12}=U_2\ket{\Psi}_{12},
\$ be valid. Then, applying \$U_2\$ to
both sides of the former relation,
exchanging the rhs and the lhs, and
utilizing the latter relation, one has:
$$U_2U_2'\ket{\Psi}_{12}=
U_2U_1'\ket{\Psi}_{12}=
U_1'U_2\ket{\Psi}_{12}=
U_1'U_1\ket{\Psi}_{12}.$$ Hence,
\$(U_1'U_1,U_2U_2')\$ are twin
unitaries, and one can define a
composition law as \$(U_1',U_2')\times
(U_1,U_2)\equiv (U_1'U_1,U_2U_2').\$
Naturally, the trivial twin unitaries
\$(1_1,1_2)\$ are the unit element.
Then the inverse of \$(U_1,U_2)\$ has
to be \$(U_1^{-1},U_2^{-1})\$, and it
is the inverse from left and from right
of the former, and it is the unique
inverse as in a group it should be. But
it is not obvious that
\$(U_1^{-1},U_2^{-1})\$ are twin
unitaries.

It is well known (and easy to see) that
the set of all (bipartite) unitaries
\$U_{12}\$ that leave the given state
\$\ket{\Psi}_{12}\$ unchanged is a
subgroup of all unitaries, the
so-called invariance group of the
vector. If \$(U_1,U_2)\$ are twin
unitaries, then \$U_1U_2^{-1}\$ leaves
\$\ket{\Psi}_{12}\$ unchanged or
envariant (cf (8a) and (8b)). Its
inverse is
\$(U_1U_2^{-1})^{-1}=U_1^{-1}(U_2^{-1})
^{-1}.\$ Then \$(U_1^{-1}, U_2^{-1})\$
are twin observables.\hfill $\Box$\\

{\bf APPENDIX C}

Those linear operators \$A\$ in a
complex separable Hilbert space are
Hilbert-Schmidt ones for which
\$\tr(A^{\dag}A)<\infty\$ (\$A^{\dag}\$
being the adjoint of \$A\$). The scalar
product in the Hilbert space of all
linear Hilbert-Schmidt operators is
\$\Big(A,B\Big)\equiv\tr(A^{\dag}B)\$
(cf the Definition after Theorem VI.21
and problem VI.48(a) in \cite{RS}).

The statement that \$\rho_n\$ converges
to \$\rho\$ in the topology determined
by the distance in the Hilbert space of
all linear Hilbert-Schmidt (HS)
operators means:
$$\lim_{n\rightarrow\infty}
||\rho-\rho_n||_{HS}^2=
\lim_{n\rightarrow\infty}
\tr(\rho-\rho_n)^2=$$
$$\lim_{n\rightarrow\infty}
\sum_k\bra{\phi_k}
(\rho-\rho_n)^2\ket{\phi_k}=0,$$ where
\$\{\ket{\phi_k}:\forall k\}\$ is an
arbitrary basis.

On the other hand, the claim that
\$\rho_n\$ converges to \$\rho\$ in the
strong operator topology means
\cite{RS} that
$$\forall \ket{\psi}:\quad
\lim_{n\rightarrow\infty}
||\rho\ket{\psi} -\rho_n\ket{\psi}||^2=
$$ $$
\lim_{n\rightarrow\infty}
\bra{\psi}(\rho-
\rho_n)^2\ket{\psi}=0.$$

Thus, the latter topology requires
convergence to zero only for each
vector separately (without any
uniformity of convergence for some
subset), whereas the former topology
requires the same uniformly for any
basis, moreover for their sum (which
may be infinite). The former topology
requires much more, and hence it is
stronger.\\

{\bf ACKNOWLEDGEMENT.} Not only through
their stimulating papers, but also by
private e-mail communication,
Schlosshauer, Barnum, Mohrhoff and
Caves helped me substantially to
understand that Zurek's argument, as
also their versions of it, is
incomplete with respect to the
probability rule; and they have
explained why it is so. I am very
grateful to them. I have obtained very
useful comments on the first draft of
this article from Zurek. I am indebted
to him. I had also some comments from
Schlosshauer and Mohrhoff. I feel
thankful to them too.

Since I have profited immensely from
the ideas of all other participants in
the "Born's rule from envariance"
enterprise, the present version is, to
a certain extent, the upshot of a
collective effort. But for all its
shortcomings and possible failures
I am the only one to blame.\\


\begin{thebibliography}{99}
\bibitem{FN2Mohrhoff}
Marcus Appleby wrote (private
communication from Mohrhoff): "Whereas
the interpretation of \QM has only been
puzzling us for about 75 years, the
interpretation of probability has been
doing so for more than 300 years."
\bibitem{Max}
The question in the title of the
subsection was raised by Schlosshauer.
He voiced the suspicion that
restriction to unitary operators might
be a way of "putting in probabilities
to get out probabilities".
\end{thebibliography}
\end{document}